\documentclass[prb,twocolumn,showpacs,preprintnumbers,amsmath,amssymb,superscriptaddress,floatfix]{revtex4}

\usepackage{graphicx}
\usepackage[T1]{fontenc}
\usepackage{lmodern}
\usepackage[latin1]{inputenc}
\usepackage{dcolumn}
\usepackage{bm}
\usepackage{color}
\usepackage[colorlinks,plainpages=false,linkcolor=blue,urlcolor=blue,citecolor=black,pdfpagemode=UseNone,pdfstartview=FitBH]{hyperref}
\makeatletter\renewcommand{\fnum@figure}[1]{\figurename~\thefigure~(color online).}\makeatother
\definecolor{orange}{rgb}{1,0.7,0}
\definecolor{magenta}{rgb}{1,0,1}

\begin{document}

\title{Strain and composition dependence of the orbital polarization in nickelate superlattices}

\author{M.~Wu}
\affiliation{Max Planck Institute for Solid State Research, Heisenbergstr.~1, 70569 Stuttgart, Germany}

\author{E.~Benckiser}
\email[]{e.benckiser@fkf.mpg.de}
\affiliation{Max Planck Institute for Solid State Research, Heisenbergstr.~1, 70569 Stuttgart, Germany}

\author{M.~W.~Haverkort}
\affiliation{Max Planck Institute for Solid State Research, Heisenbergstr.~1, 70569 Stuttgart, Germany}
\affiliation{Quantum Matter Institute, University of British Columbia, Vancouver, B.C. V6T 1Z1, Canada}

\author{A.~Frano}
\affiliation{Max Planck Institute for Solid State Research, Heisenbergstr.~1, 70569 Stuttgart, Germany}
\affiliation{Helmholtz-Zentrum Berlin f\"{u}r Materialien und Energie, Wilhelm-Conrad-R\"{o}ntgen-Campus BESSY II, Albert-Einstein-Str. 15, D-12489 Berlin, Germany}

\author{Y.~Lu}
\affiliation{Max Planck Institute for Solid State Research, Heisenbergstr.~1, 70569 Stuttgart, Germany}

\author{U.~Nwankwo}
\affiliation{Max Planck Institute for Solid State Research, Heisenbergstr.~1, 70569 Stuttgart, Germany}

\author{S.~Br\"{u}ck}
\affiliation{Max Planck Institute for Intelligent Systems, Heisenbergstr.~3, 70569 Stuttgart, Germany}
\affiliation{Physikalisches Institut and Roentgen Research Center for Complex Material Systems, Universit\"{a}t W\"{u}rzburg, 97074 W\"{u}rzburg, Germany}

\author{P.~Audehm}
\affiliation{Max Planck Institute for Intelligent Systems, Heisenbergstr.~3, 70569 Stuttgart, Germany}

\author{E.~Goering}
\affiliation{Max Planck Institute for Intelligent Systems, Heisenbergstr.~3, 70569 Stuttgart, Germany}

\author{S.~Macke}
\affiliation{Quantum Matter Institute, University of British Columbia, Vancouver, B.C. V6T 1Z1, Canada}

\author{V.~Hinkov}
\affiliation{Quantum Matter Institute, University of British Columbia, Vancouver, B.C. V6T 1Z1, Canada}

\author{P.~Wochner}
\affiliation{Max Planck Institute for Intelligent Systems, Heisenbergstr.~3, 70569 Stuttgart, Germany}

\author{G.~Christiani}
\affiliation{Max Planck Institute for Solid State Research, Heisenbergstr.~1, 70569 Stuttgart, Germany}

\author{S.~Heinze}
\affiliation{Max Planck Institute for Solid State Research, Heisenbergstr.~1, 70569 Stuttgart, Germany}

\author{G.~Logvenov}
\affiliation{Max Planck Institute for Solid State Research, Heisenbergstr.~1, 70569 Stuttgart, Germany}

\author{H.-U.~Habermeier}
\affiliation{Max Planck Institute for Solid State Research, Heisenbergstr.~1, 70569 Stuttgart, Germany}

\author{B.~Keimer}
\email[]{b.keimer@fkf.mpg.de}
\affiliation{Max Planck Institute for Solid State Research, Heisenbergstr.~1, 70569 Stuttgart, Germany}

\date{\today}

\begin{abstract}
A combined analysis of x-ray absorption and resonant reflectivity data was used to obtain the orbital polarization profiles of superlattices composed of four-unit-cell-thick layers of metallic LaNiO$_3$ and layers of insulating \textit{R}XO$_3$ (\textit{R}\,=\,La, Gd, Dy and X\,=\,Al, Ga, Sc), grown on substrates that impose either compressive or tensile strain. This superlattice geometry allowed us to partly separate the influence of epitaxial strain from interfacial effects controlled by the chemical composition of the insulating blocking layers. Our quantitative analysis reveal orbital polarizations up to 25\,$\%$. We further show that strain is the most effective control parameter, whereas the influence of the chemical composition of the blocking layers is comparatively small.
\end{abstract}

\pacs{73.21.Cd, 78.70.-g, 61.05.cm, 73.20.-r, 71.20.Be}

\maketitle
\section{Introduction}
The prospect of designing new superconductors by orbital engineering\cite{Chaloupka2008} has recently triggered intense research activity on artificial superlattices of the metallic perovskite LaNiO$_3$ (LNO) and isostructural band insulators. The Ni$^{3+}$ ions in bulk LNO adopt the electron configuration $t_{2g}^6 e_g^1$, and initial analytical calculations\cite{Chaloupka2008} indicated a single Fermi surface with dominant $d_{x^2-y^2}$ character for the $e_g$ electrons in suitably prepared nickelate superlattices. The shape of this Fermi surface was predicted to resemble the one of the cuprate superconductors, raising hopes for ``engineered'' high-temperature superconductivity. Subsequent calculations based on density-functional theory (DFT) addressed the influence of confinement, strain, structural distortions, chemical composition of the insulating layers, and electronic correlations on the orbital polarization of the Ni $e_g$ electron, with widely divergent results.\cite{Hansmann2009,Han2010,Hansmann2010,Xiaoping2011,Blanca2011,Han2011,Han2012} Whereas some DFT calculations supported original predictions,\cite{Han2010,Hansmann2010,Xiaoping2011} Han \textit{et al.}\cite{Han2011} later reported that the combined effect of the on-site Hund interaction and the covalency of the nickel-oxygen bond greatly reduces the orbital polarization, so that the orbital degeneracy retains its dominant influence on the electronic structure of the nickelates even under the most favorable conditions. This conclusion received support from x-ray absorption spectroscopy (XAS) studies of ultrathin LNO films\cite{Chakhalian2011} and superlattices.\cite{Freeland2011} Whereas films under compressive strain showed a slightly enhanced occupation of the $d_{3z^2-r^2}$ orbital, XAS data for films under tensile strain were interpreted as evidence of a charge-ordering instability with negligible orbital polarization.\cite{Chakhalian2011} If confirmed, this strongly asymmetric orbital-lattice coupling would have major implications for the design of ``orbitally engineered'' oxide heterostructures in general.

This article describes the results of an experimental program designed to quantitatively determine the orbital occupation of nickelate superlattices, to test the predictions of the DFT calculations, and to explore the feasibility of phase control by orbital engineering. We present results from resonant x-ray reflectivity and x-ray linear dichroism measurements on a series of (4\,u.c.//4\,u.c.)$\times 8$ superlattices composed of four consecutive pseudocubic unit cells (u.c.) of LNO and equally thick layer stacks of the band insulators LaAlO$_3$ (LAO), LaGaO$_3$ (LGO), DyScO$_3$ (DSO) and GdScO$_3$ (GSO). The superlattices were grown by pulsed-laser deposition on substrates with different signs and magnitudes of the lattice mismatch to bulk LNO, i.e. YAlO$_3$ (YAO), LaSrAlO$_4$ (LSAO), SrTiO$_3$ (STO), DSO, and GSO with $a_{\rm YAO}$\,$<$\,$a_{\rm LSAO}$\,$<$\,$a_{\rm LNO\, bulk}$\,$<$\,$a_{\rm STO}$\,$<$\,$a_{\rm DSO}$\,$<$\,$a_{\rm GSO}$ where $a$ is the lattice constant of the perovskite structure. Orbital polarizations are quantified through the application of sum rules to the transition-metal $L$-edge XAS measured with linearly polarized soft x-rays. The ``orbital reflectometry'' technique introduced in Ref.~\onlinecite{Benckiser2011} allowed us to obtain quantitative layer-resolved orbital polarization profiles within the LNO layer stacks, and to partly disentangle the orbital polarization originating from the tetragonal distortion induced by the substrate and the spatial confinement (which affects all four LNO layers in the stack) from the the change in chemical composition across the LNO-\textit{R}XO interface (which largely affects the interfacial layers). In contrast to the conclusions from the previous experiments,\cite{Chakhalian2011,Freeland2011} our results indicate a linear orbital-lattice coupling, and confirm the stabilization of the planar $d_{x^2-y^2}$ orbital under tensile strain. Going beyond this experimental work, we specify the layer-resolved orbital polarization quantitatively and directly compare the experimental data with the results of the DFT calculations.

\section{Experimental Details}

\begin{figure*}[tb]
\center\includegraphics[width=0.85\textwidth]{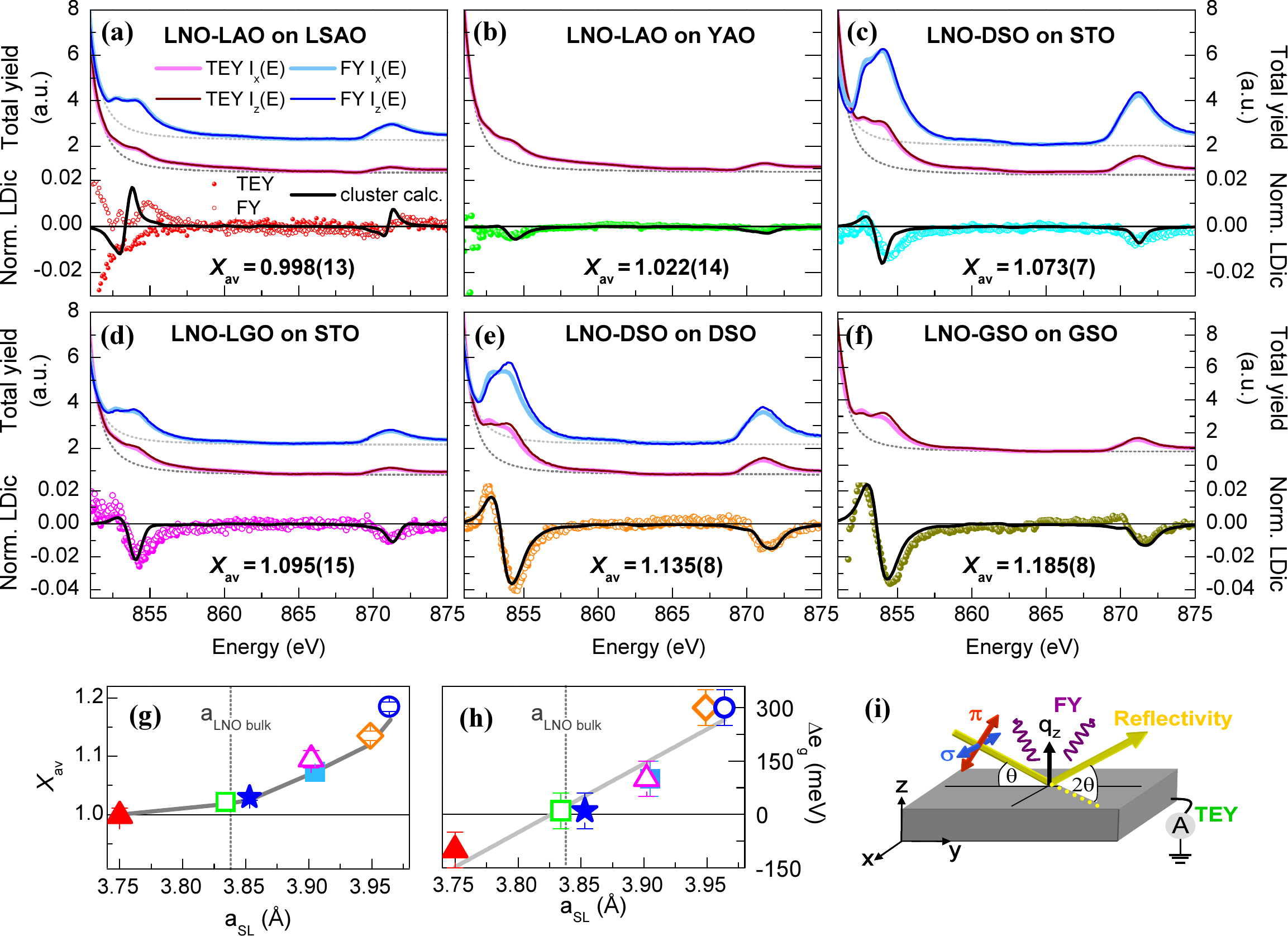}
\caption{(a)-(f) XAS spectra (FY shifted by +1.5 for clarity) measured with linearly polarized light. Dotted grey lines show the results of Lorentzian fits to the tail of the La $M_4$ lines. The normalized difference spectra ($I_x(E)$-$I_z(E)$)/($\frac{1}{3}$(2$I_x$ +$I_z$)) are shown directly below the corresponding spectra, together with the results of the cluster calculation. (g) Hole ratio $X_{\rm av}$ obtained via the sum rule (Eq.~\ref{eq:sum rule}) and (h) crystal field splitting $\Delta e_g$ obtained from the cluster calculation versus the in-plane lattice constant $a_{\rm SL}$ (see the Appendix) of LNO-LAO on LSAO (\textcolor{red}{\large{$\blacktriangle$}}), LNO-LAO on YAO (\textcolor{green}{\large{$\square$}}), LNO-LAO on STO (Ref.~\onlinecite{Benckiser2011}; \textcolor{blue}{\large{$\bigstar$}}), LNO-DSO on STO (\textcolor{cyan}{\large{$\blacksquare$}}), LNO-LGO on STO (\textcolor{magenta}{\large{$\triangle$}}), LNO-DSO on DSO (\textcolor{orange}{\LARGE{$\diamond$}}), and LNO-GSO on GSO (\textcolor{blue}{\LARGE{$\circ$}}). (i) Sketch of the measurement geometry.}
\label{Fig:LDIC_all}
\end{figure*}

LNO-\textit{R}XO superlattices were grown by pulsed laser deposition and characterized by atomic force microscopy, dc resistivity, transmission electron microscopy,\cite{Detemple2011,Detemple2012,DetempleThesis} and hard x-ray diffraction (see the Appendix for details).

The resonant x-ray reflectivity and x-ray absorption spectroscopy (XAS) measurements were performed at the UE56/2-PGM1 soft-x-ray beam line at BESSY II in Berlin, Germany, using the advanced three-axis ultrahigh-vacuum reflectometer described in Ref.~\onlinecite{Brueck2008}. A sketch of the measurement geometry is shown in Fig.~\ref{Fig:LDIC_all}i. Linearly polarized soft x-rays ($\sigma$ and $\pi$-polarization) tuned to the Ni $L$ edge were used to measure XAS spectra in two collection modes, total electron yield (TEY) and total fluorescence yield (FY). Reflected intensities were detected with a diode. All intensities were normalized to the incoming intensity measured with a gold mesh. Since our dc resistivity measurements and a prior study showed that superlattices with 4\,u.c.-thick LNO layer stacks remain metallic at all temperatures and do not exhibit any temperature-induced changes of the electronic and magnetic structure,\cite{Boris2011,Frano2012} we present room-temperature data.

\section{Results and Discussion}

\subsection{Spatially averaged orbital polarization}

We first discuss the XAS data shown in Fig.~\ref{Fig:LDIC_all}. Except for the LNO-LAO on LSAO superlattice (Fig.~\ref{Fig:LDIC_all}a), the spectra of all superlattices clearly show a polarization dependence at the Ni $L$-edge, which we attribute to natural linear dichroism. The magnitude of the observed dichroism varies substantially between superlattices of different composition and can be clearly seen in the normalized difference spectra (lower panels in Fig.~\ref{Fig:LDIC_all}a-f). In particular we point out that the observed dichroism in LNO-\textit{R}ScO superlattices is substantial, having in mind that even in the case of full $x^2 - y^2$ orbital polarization in the atomic limit the integrated intensity of the spectrum for $x$ polarization is about 60$\%$ of that of the $z$ polarization. Although the spectra obtained in TEY and FY detection modes differ in spectral weight and line shape, their polarization dependencies agree remarkably well (lower panels in Fig.~\ref{Fig:LDIC_all}a-f). This confirms that the observed linear dichroism is robust and not related to surface effects.

In order to quantitatively analyze the observed dichroism, we applied the sum rule for linear dichroism,\cite{vanderLaan1994,Benckiser2011} which relates the ratio of holes in the Ni $e_g$ orbitals to the energy-integrated XAS intensities across the Ni $L$ edge $I_{x,z}=\int_{L_{3,2}}I_{x,z}(E)dE$ for in-plane ($x$) and out-of-plane ($z$) polarization, respectively:
\begin{equation}\label{eq:sum rule}
X=\frac{h_{3z^2-r^2}}{h_{x^2-y^2}}={\left(\frac{3I_{z}}{4I_{x}-I_{z}}\right)}.
\end{equation}
Here $h_{x^2-y^2}$ and $h_{3z^2-r^2}$ are the hole occupation numbers of orbitals with $x^2$-$y^2$ and $3z^2$-$r^2$ symmetry.

Before proceeding to describe the analysis of the linear dichroism, we add a remark about data processing. Since the La $M_4$ line partially overlaps the Ni $L_3$ contribution, it has to be subtracted before integrating the Ni XAS spectra. We carefully estimated the error associated with this subtraction. The results presented in the following were obtained by subtracting Lorentzian line shapes from the TEY and FY data (dashed lines in Fig.~\ref{Fig:LDIC_all}). Within our error bars, we obtained identical results when subtracting a La $M$-edge reference spectrum measured on LaCoO$_3$. Because there is a substantial difference in the La $M$-edge line shape measured in TEY and FY, different Lorentzians were subtracted from these spectra. Note, however, that there is no linear dichroism at the La $M$-edge, so that identical Lorentzians can be subtracted for light polarization parallel to $x$ and $z$. To further crosscheck our results, we compared the sum rule results obtained by integrating across the Ni $L_{3,2}$ lines with those obtained by integrating only across the Ni $L_2$ region, which is not affected by the La $M_4$-line subtraction. Again we found that both results are identical within the given error bars.

In this way, the spatially averaged hole ratio $X_{\rm av}$ was calculated from the average of values determined from the sum rule analysis of TEY and FY XAS data. This quantity is shown in each panel of Fig.~\ref{Fig:LDIC_all}a-f. In Fig.~\ref{Fig:LDIC_all}g we show $X_{\rm av}$ as a function of the measured in-plane lattice parameter $a_{\rm SL}$ (see the Appendix). $X_{\rm av}$ increases monotonically with increasing $a_{\rm SL}$. We will further discuss this relationship below.

XAS is a well-established technique for studying the unoccupied site- and symmetry-projected electronic density of states of solids, providing the possibility to compare experimental results with single-particle band-structure calculations, often obtained using DFT. In the past it was demonstrated that for a satisfactory description of the observed fine structure of transition-metal $L$-edges, it is important to include many-body effects, including in particular the interaction of the $2p$ core hole created in the absorption process with electrons in the partially filled $3d$ final state. This can be done by atomic multiplet theory,\cite{deGroot1990} an approach which we discuss in more detail in the next paragraph on the basis of the cluster calculation results. At this point, we emphasize that for the determination of orbital polarizations from \textit{energy-integrals} of the XAS spectra across the full Ni $L$-edge via Eq.~\ref{eq:sum rule}, a detailed understanding of the XAS fine structure is not necessary. While the effect of the core hole potential enters the Hamiltonian of the system, the sum rule is independent of it, and therefore $X_{\rm av}$ reflects the polarization-dependence of the $d$-projected unoccupied density of states.

\begin{figure}[t]
\center\includegraphics[width=0.99\columnwidth]{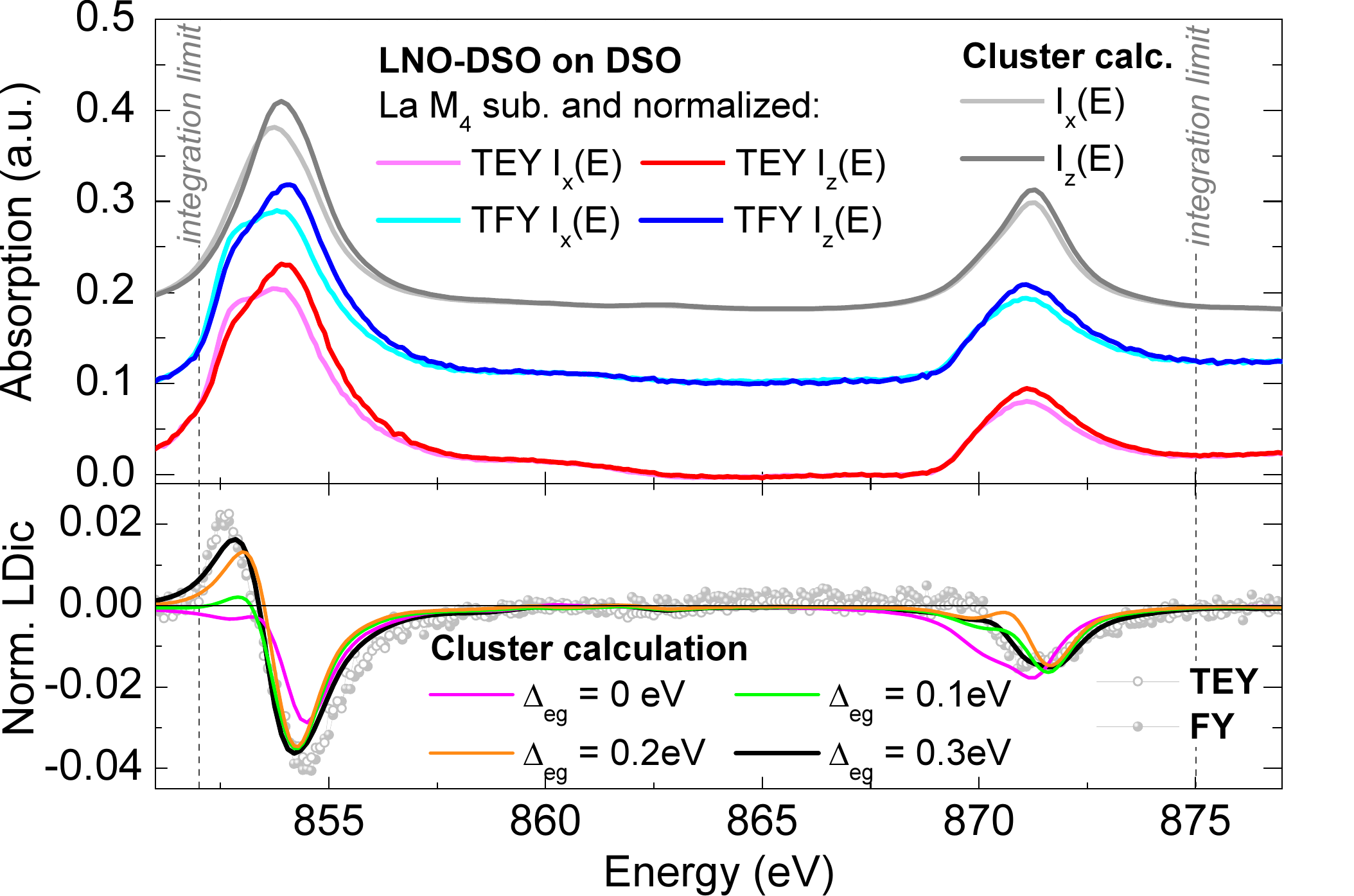}
\caption{Top panel: Polarization-dependent XAS spectra (TEY and FY) after subtraction of the La $M_4$ line (Lorentzian fit) together with the spectra obtained from our cluster calculation ($\gamma = 0.6$, $\delta = 0.4$ and  $\Delta e_g = 300$\,meV). All spectra are normalized by their polarization-averaged integral (($A = 2I_x + I_z)/3$). Bottom panel: Normalized difference spectra $(I_x(E)-I_z(E))/A$. The results of the cluster calculation are shown for $\gamma = 0.6$, $\delta = 0.4$ and different values of $\Delta e_g$, ranging from 0 to 300\,meV.}
\label{Fig:ClusterCalc}
\end{figure}

As a further step in the analysis of the spectroscopic data, we performed a cluster calculation for Ni $3d^7$ in a tetragonal ligand field, in which both the orbital polarization and the energy difference between the Ni $e_g$ orbitals, $\Delta e_g$, were adjusted to maximize agreement with the dichroic difference spectra. Additional parameters are radial integrals, Slater integrals, and spin-orbit coupling constants, which were obtained by atomic Hartree-Fock calculations, as well as 10Dq\,=\,2\,eV, the hybridization strength $pd\sigma$=-1.88, $pd\pi = - pd\sigma /2.17$, and the charge-transfer energy, which was estimated by $\Delta_{3+} = \Delta_{2+} - U_{dd} = -0.5$\,eV with $U_{dd}=7$\,eV and $\Delta_{2+} = 6.5$\,eV from Ni$^{2+}$.\cite{Tanaka1994,Schuessler2005} The measured spectra were then described as linear superpositions of spectra $I_{x,z}^{x^2-y^2}(E)$, $I_{x,z}^{3z^2-r^2}(E)$ calculated for 100\,$\%$ occupation of either the $x^2-y^2$ or the $3z^2-r^2$ orbital and for polarization of the incoming light parallel to the $x$ and $z$ direction, respectively: $I_{x,z}(E) = \gamma I_{x,z}^{x^2-y^2}(E) + \delta I_{x,z}^{3z^2-r^2}(E)$. The admixture coefficients ($\gamma$ and $\delta$ with $\gamma + \delta = 1$) and $\Delta e_g$ were then varied until the best agreement with the experimental linear-dichroic difference spectra was obtained (Fig.~\ref{Fig:ClusterCalc}). Since the cluster calculation is a local approach and the LNO layers in all superlattices studied here are metallic (as demonstrated by dc transport and optical spectroscopy\cite{Boris2011}), it is not surprising that the lineshapes of the spectra for light polarization parallel to $x$ and $z$ are not exactly reproduced (see Fig.~\ref{Fig:ClusterCalc} and the discussion in Ref.~\onlinecite{Benckiser2011}). We emphasize, however, that the dichroic difference spectra ($I_x(E) -I_z(E)$) are almost independent of the individual lineshapes and describe our experimental data very well (Fig.~\ref{Fig:ClusterCalc}). While a preferred orbital occupation of one of the $e_g$ orbitals is seen as an intensity difference between $I_x(E)$ and $I_z(E)$ spectra, the effect of the crystal-field splitting manifests itself as an energy shift between these spectra, which results in a derivative-like lineshape of the difference spectra (lower panels in Fig.~\ref{Fig:LDIC_all}a-f). The variation of $\Delta e_g$ as a function of the measured in-plane lattice constant $a_{\rm SL}$ (Fig.~\ref{Fig:LDIC_all}h) is consistent with the behavior of $X_{\rm av}$ (Fig.~\ref{Fig:LDIC_all}g). Whereas the value for the LNO-LAO superlattice under compressive strain ($\Delta e_g \approx -100$ meV) agrees with the one reported earlier for a similar sample,\cite{Chakhalian2011,Freeland2011} we see a comparable shift also for tensile strain with a roughly linear dependence of $\Delta e_g$ on $a_{\rm SL}$ based on a large number of samples. (See Section~\ref{subsec:Layer-resolved orbital polarization} for a discussion of the deviation of the $X_{\rm av}$-versus-$a_{\rm SL}$ relation from linearity.) Our results clearly indicate an approximately linear orbital-lattice coupling, and confirm the stabilization of the planar $d_{x^2-y^2}$ orbital under tensile strain. This result differs from the previously reported asymmetry between the behavior under tensile and compressive strain of ultrathin LaNiO$_3$ thin film\cite{Chakhalian2011} and LaNiO$_3$ based superlattices,\cite{Freeland2011} at least in the (4\,u.c.//4\,u.c.) superlattice structures investigated here.

\begin{figure}[b]
\includegraphics[width=0.99\columnwidth]{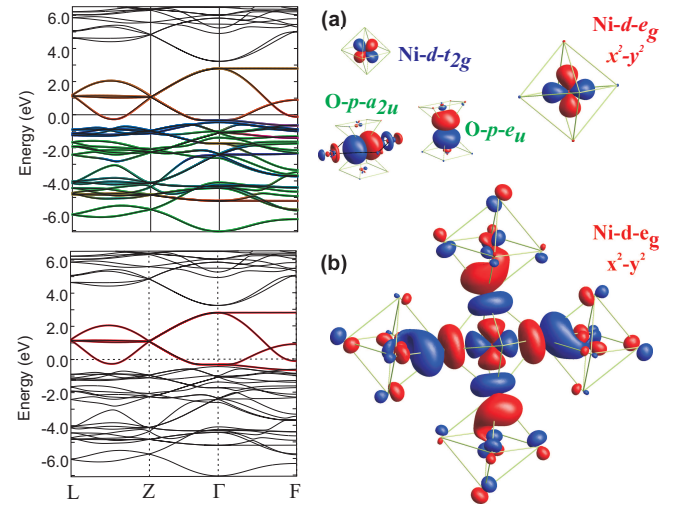}
\caption{Band structure (left) and Wannier orbitals (right) of bulk LaNiO$_3$ (space group R$\bar{3}$c, Ref.~\onlinecite{Munoz1992}): (a) down-folded
to atomic-like Ni-$d$ and O-$p$ orbitals and (b) Wannier orbitals obtained by downfolding the antibonding Ni-$d$ O-$p$ bands to extended Ni-d$_{e_g}$ orbitals, explicitly including covalency. The difference in phase of the wave functions is depicted by red and blue colors. The color coding for the band structure is as follows: red corresponds to Ni $e_g$ , blue to Ni $t_{2g}$ and green to O $p$ character of the bands. The calculations were performed using the Stuttgart-NMTO code.\cite{Andersen2000,Andersen2000b}}
\label{Fig:Orbitals}
\end{figure}

In order to compare our experimental results with the DFT predictions mentioned above, we define the orbital polarization following Refs.~\onlinecite{Han2010,Han2011} and using Eq.~\ref{eq:sum rule}:
\begin{equation}
P=\frac{n_{x^2-y^2}-n_{3z^2-r^2}}{n_{x^2-y^2}+n_{3z^2-r^2}}=\left(\frac{4}{n_{e_g}}-1\right)\frac{(X-1)}{(X+1)},
\label{Eq:OrbPol}
\end{equation}
where $n_{x^2-y^2}=2-h_{x^2-y^2}$ and $n_{3z^2-r^2}=2-h_{3z^2-r^2}$ denote the numbers of electrons in orbitals of $x^2$-$y^2$ and $3z^2$-$r^2$ symmetry, and $n_{e_g}=4-h_{e_g}$ is their sum. The latter parameter deserves particular attention, since hybridization between Ni $d$ and O $p$ is not negligible and can be affected by strain and the composition of the insulating material. In order to calculate $P$ of the local, atomic-like Ni $d$ orbitals for all different superlattices, the value of $n^{\rm atomic}_{e_g}$ has to be known. Theoretical values show fairly large variations of $n^{\rm atomic}_{e_g}$\,=\,$1.5 - 2.1$ as a function of composition and interactions,\cite{Han2010,Han2013} but an experimental determination is a difficult task. Here we suggest to consider the orbital polarization of extended Wannier orbitals in order to provide a well-defined quantitative description for the discussion and comparison of superlattices with possibly different hybridization. The orbital polarization of those extended Wannier orbitals is obtained via Eq.~\ref{Eq:OrbPol} using $n_{e_g} = 1$ for all superlattices studied. To illustrate the different wave functions, we performed DFT calculations using the experimentally reported crystal structure of bulk LaNiO$_3$\cite{Munoz1992} (for details see Ref.~\onlinecite{Haverkort2012}). We considered two cases: (i) a large basis of five atomic-like Ni $d$ and three O $p$ orbitals with $n^{\rm atomic}_{e_g}=1.8$ due to hybridization (top panel in Fig.~\ref{Fig:Orbitals}) and (ii) a small basis of extended Wannier orbitals, labeled with d, representing the antibonding Ni $e_g$ and O $p$ states near the Fermi level with $n_{e_g} = 1$ (bottom panel in Fig.~\ref{Fig:Orbitals}). The latter orbitals are very close to the band eigenstates and reflect the covalency due to their large weight at the oxygen positions, but exhibit the full symmetry of the $e_g$ orbitals. Describing our results using this basis functions does not require the knowledge of the strain and composition dependent values $n^{\rm atomic}_{e_g}$, since the differences in hybridization are reflected in a local change of the Wannier functions. Furthermore, a similar orbital basis set was used to calculate orbital polarizations in Refs.~\onlinecite{Hansmann2009,Han2010,Han2011}, i.e.\ only bands spanning a small energy window close to the Fermi level were integrated to obtain the corresponding occupation numbers $n_{x^2-y^2}$ and $n_{3z^2-r^2}$ (-3 and -1.5\,eV to E$_F$=0 in Ref.~\onlinecite{Han2011} and \onlinecite{Han2010}, respectively).\footnote{Detailed inspection of the revised results\cite{Han2013} of Ref.~\onlinecite{Han2011} indicates that the Fermi surface properties, in particular the size of the central Fermi surface patch, is reflected in $P_{-3\rightarrow 0}$ (extended Wannier orbitals) rather than in $P_{-\infty\rightarrow 0}$ (atomic-like orbitals). While the values of $P_{-3\rightarrow 0}$ are materially different for the interacting cases (23-37$\%$, large central patch) compared to the non-interacting case (50$\%$; very small central patch), the values for $P_{-\infty\rightarrow 0}$ fall into a fairly narrow range of 11-17$\%$ for \textit{all} cases.} In particular the total number of these states is $n_{e_g}\approx 1$ and therefore comparable to our experimental results on an absolute scale. For all superlattices we obtain positive values in the range $P_{\rm av} = 0 - 25 \%$ for the spatially averaged orbital polarization, corresponding to a substantially enhanced occupancy of the orbital with $x^2-y^2$ symmetry.

Even without further analysis, we can thus conclude that orbital engineering of nickelates is a much more potent tool than indicated by recent first principles calculations,\cite{Han2011} where $P$ did not exceed 9$\%$ even under the most favorable conditions. In Ref.~\onlinecite{Han2011} the system with the largest orbital polarization found in Ref.~\onlinecite{Han2010} was considered, i.e.\ a LaNiO$_3$/LaInO$_3$ superlattice with (1\,u.c.//1\,u.c.) structure and in-plane lattice parameters set to the values of SrTiO$_3$ ($a=3.90$\AA). For nonzero electron-electron interaction strength (Hubbard $U \neq 0$) orbital polarizations $P_{-3\rightarrow 0}=5-9\,\%$ were obtained by integrating the density of states in the range -3\,eV to 0. As argued above, calculating $P$ via Eq.~\ref{Eq:OrbPol} using $n_{e_g}=1$ yields a good estimate of $P_{-3\rightarrow 0}$. Although we did not study the particular superlattice considered in Ref.~\onlinecite{Han2011}, we found $P_{\rm}=14$\,$\%$ for the LNO-LGO on STO superlattice with (4\,u.c.//4\,u.c.) structure, exceeding the predicted values by almost a factor of two, even under less favorable conditions. We note that the comparison with our experimental results prompted the publication of an Erratum\cite{Han2013} with revised values of $P_{-3\rightarrow 0}=23-37\,\%$ for $U\neq 0$.\footnote{When integrating over the entire bandwidth values of $P_{-\infty\rightarrow 0}=11\,\%$ and $n^{\rm atomic}_{e_g}=1.52$ were obtained in the most favored case (d) ($U=6$\,eV) of Refs.~\onlinecite{Han2011,Han2013}. Using the same value $n^{\rm atomic}_{e_g}=1.52$ in Eq.~\ref{Eq:OrbPol} for the LNO-LGO superlattice we obtain an estimate of the orbital polarization of the atomic-like orbitals: $P_{\rm atomic}\approx 8\,\%$.} These values are in much better agreement with our results.

\subsection{Layer-resolved orbital polarization}\label{subsec:Layer-resolved orbital polarization}

\begin{figure}[tb]
\includegraphics[width=0.99\columnwidth]{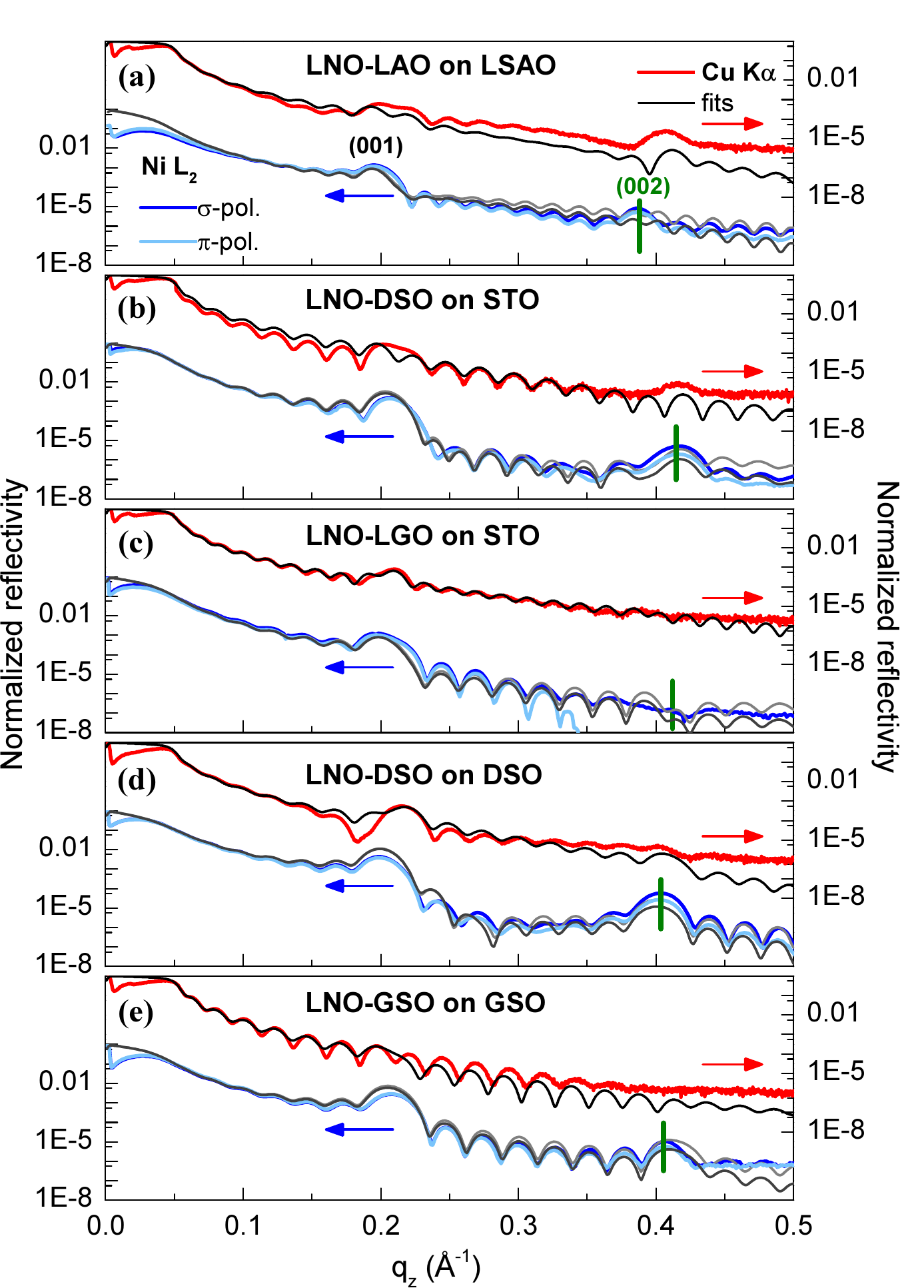}
\caption{Reflectivity as a function of $q_z$ for the (a) LNO-LAO on LSAO, (b) LNO-DSO on STO, (c) LNO-LGO on STO, (d) LNO-DSO on DSO, and (e) LNO-GSO on GSO superlattice. The $q_z$ values at (002), chosen for the constant-$q_z$ shown in Fig.~3 of the main text, are marked by green vertical lines and correspond to values of (a) 0.3880\,\AA$^{-1}$, (b) 0.4146\,\AA$^{-1}$, (c) 0.4120\,\AA$^{-1}$, (d) 0.4035\,\AA$^{-1}$, (e) 0.4055\,\AA$^{-1}$.}
\label{Fig:Refl}
\end{figure}

\begin{figure}[t]
\center\includegraphics[width=0.88\columnwidth]{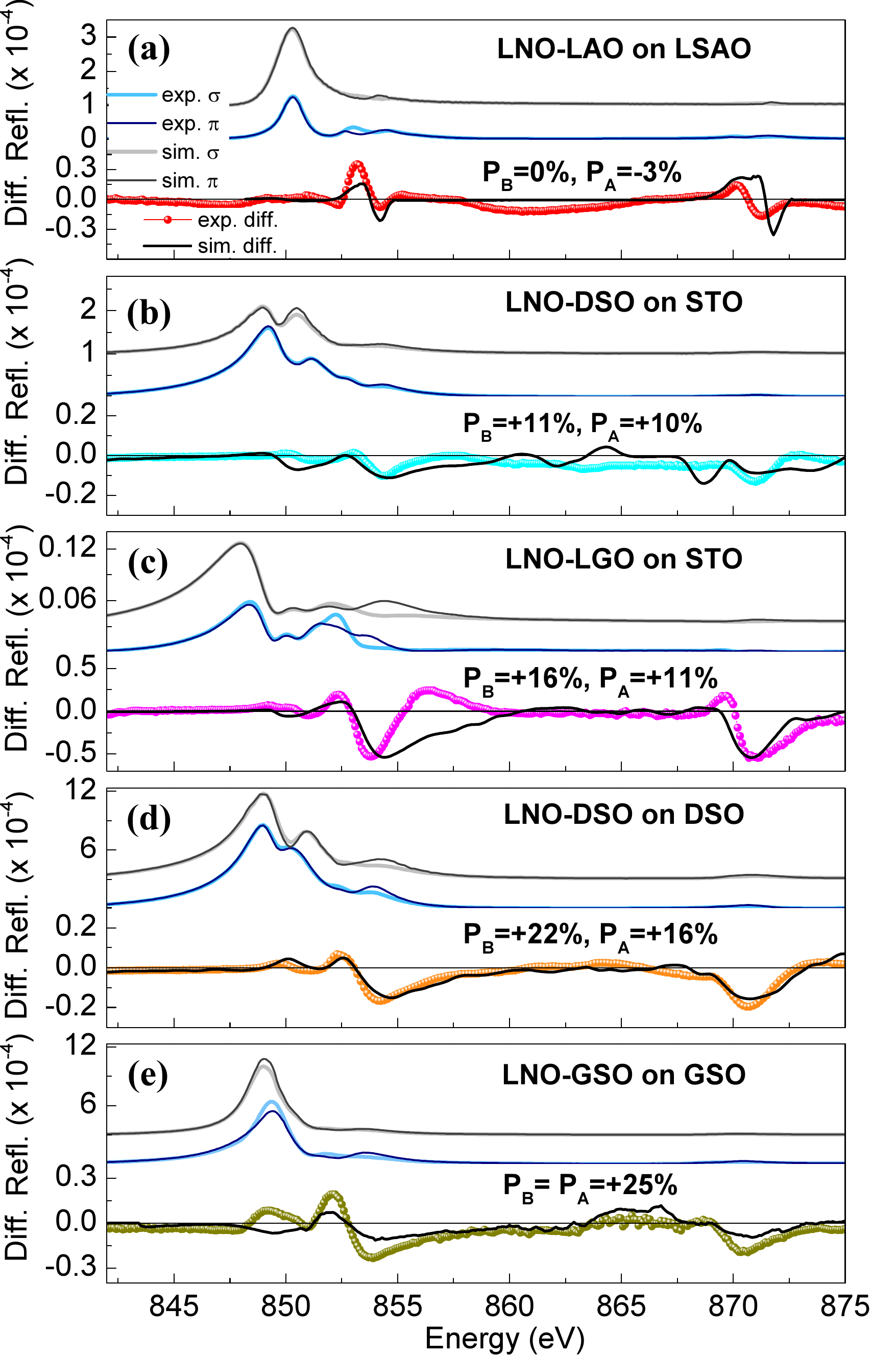}
\caption{Experimental and simulated constant-$q_z$ energy scans at the (002) superlattice peak of (a) LNO-LAO on LSAO, (b) LNO-DSO on STO, (c) LNO-LGO on STO, (d) LNO-DSO on DSO, and (e) LNO-GSO on GSO. The experimentally obtained normalized difference ($I_{\sigma}(E)$-$I_{\pi}(E)$)/($I_{\sigma}(E)$+$I_{\pi}(E)$) is shown directly below the corresponding spectrum together with the simulated one. The obtained layer-resolved orbital polarizations within the LNO layer stack, $P_B$ (interface layer) and $P_A$ (inner layers), are stated in each panel.}
\label{Fig:ConstQz_all}
\end{figure}

In an effort to elucidate the influences of strain and interfacial chemistry on the orbital polarization, we have determined layer-resolved profiles of $P$. For this purpose, the intensity of the specularly reflected beam was measured as a function of momentum transfer ($q_z$) and energy (Fig.~\ref{Fig:Refl} and \ref{Fig:ConstQz_all}).

In the first step of our analysis, a structural model was obtained from fits of models with 9 independent parameters (thickness and roughness of the individual layerstacks) to the $q_z$-dependent non-resonant reflectivity using the reflectivity fitting tool ReMagX.\cite{ReMagX} To improve the fits we allowed the layer directly adjacent to the substrate and the top layer at the surface to have different thickness and roughness. Within the error, for all superlattices our fits confirm the expected (4\,u.c.//4\,u.c.) structure with interface roughnesses around 1\,u.c (see Tab.~\ref{Tab:Superlattices}). In the following, the structural parameters were fixed for the simulation of the dichroic reflectivity. In order to account for the strong changes in optical constants across the resonances we scaled the measured linearly polarized XAS data to the theoretical values.\cite{Chantler2000} While for all other superlattices we used the measured linearly polarized XAS data to construct the optical constants, for the LNO-LAO on LSAO superlattice these data were not available since $P_{\rm av}\sim$\,0. For the inner layers with $P_A$\,=\,-3$\%$ we calculated the dichroic difference (LDic) from our cluster calculation with $\Delta e_g = -100$\,meV and admixture coefficients $\gamma = 0.485$ and $\delta = 0.515$. In order to construct the optical constants, we added this difference to the averaged experimental XAS data, i.e.\ XAS + $\frac{1}{2}$LDic for $x$ and XAS-$\frac{1}{2}$LDic for $z$ polarized light, respectively. The results of fits to the non-resonant ($E=8048$\,eV) and resonant (Ni $L_2$) $q_z$-dependent reflectivity data are shown in Fig.~\ref{Fig:Refl}.

Figure~\ref{Fig:ConstQz_all} shows the energy- and polarization-dependent resonant reflectivity of various samples with fixed momentum transfer $q_z$ close to the (002) superlattice peak, at which the scattering intensity reflected from symmetric superlattices is sensitive to modulations within the LNO layer stack.\cite{Benckiser2011}$^{,}$\footnote{Note that the actual $q_z$ values given in the caption of Fig.~\ref{Fig:Refl} vary slightly for the different superlattices because of their (small) structural differences.} Using the numerical routines\cite{ReMagX} introduced in Ref.~\onlinecite{Benckiser2011}, we computed the normalized dichroic difference spectra for models with different orbital polarization in the inner ($P_A$) and outer ($P_B$) LNO layers of the four-layer stack (bottom panels of Fig.~\ref{Fig:ConstQz_all}), with $P_{\rm av}$ kept fixed at the value determined from XAS.

For superlattices grown on substrates inducing tensile strain ($a_{\rm SL} > a_{\rm LNO\, bulk}$), good agreement between the measured and simulated spectra was obtained only for models with positive $P_A$ and $P_B$, corresponding to enhanced $x^2-y^2$ occupancy in all LNO layers. The LNO-LAO superlattice on LSAO with $a_{\rm SL} < a_{\rm LNO\, bulk}$ is a special case, because the XAS spectrum shows that $P_{\rm av} \sim 0$. This, however, does not imply that the orbital polarization vanishes. In fact, the best agreement with the reflectivity data was obtained on the basis of a model with $P_B \sim 0$ and $P_A \sim -3 \%$ (Fig.~\ref{Fig:ConstQz_all}a), a result that could not have been obtained if only XAS data had been available.\footnote{Note that the averaged orbital polarization $0.5(P_A+P_B)=-1.5$\,$\%$ is within our experimental error of $P_{av}$\,=\,0$\pm$\,2\,$\%$ obtained from XAS.}

\begin{figure}[tb]
\includegraphics[width=0.99\columnwidth]{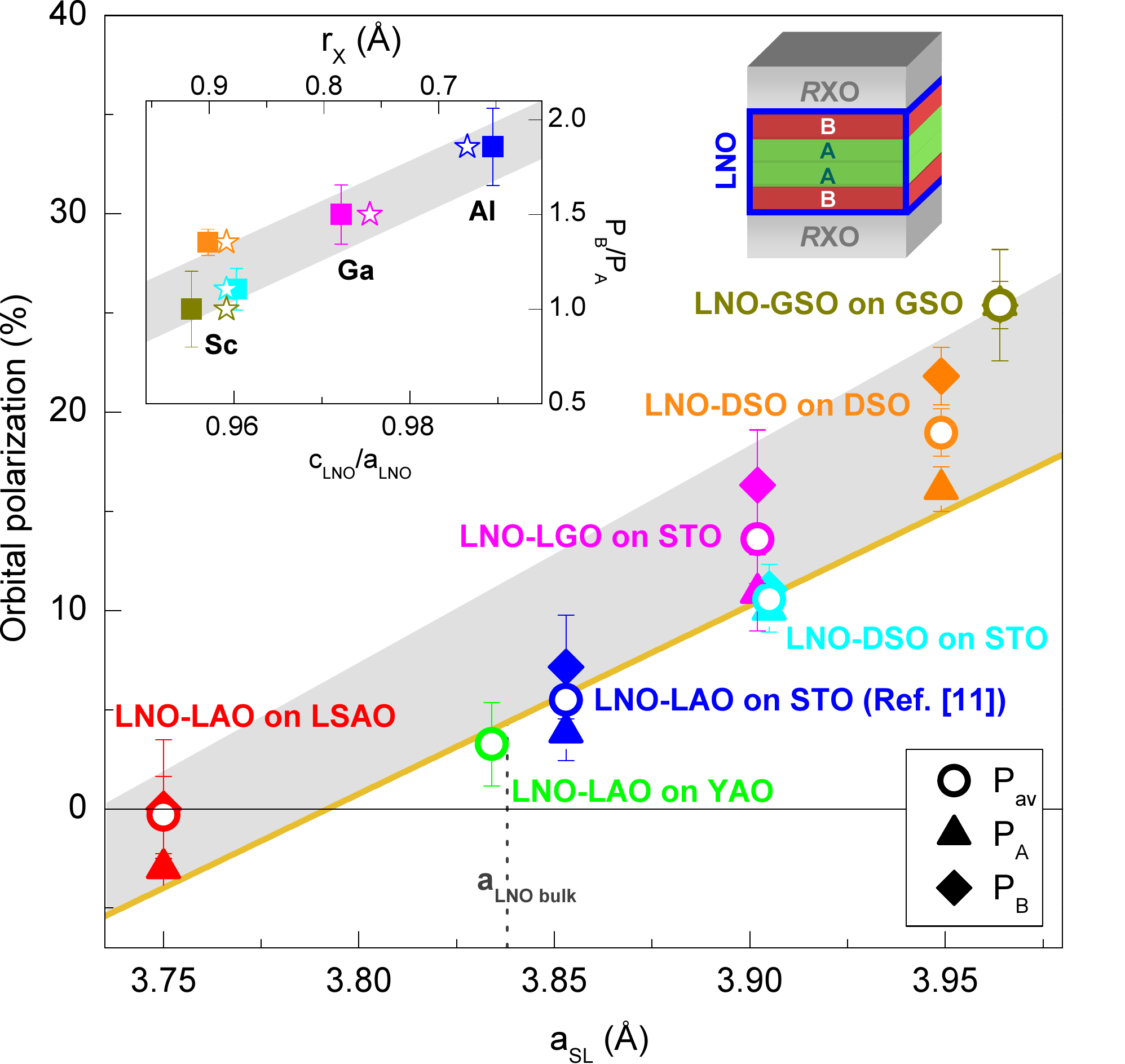}
\caption{Averaged ($P_{\rm av}$) and layer-resolved ($P_A$, $P_B$) orbital polarizations obtained from the combined analysis of XAS and reflectivity as a function of the in-plane lattice constant $a_{\rm SL}$ measured by x-ray diffraction. Inset: ratio $P_B/P_A$ vs.\ the lattice constant ratio $c_{\rm LNO}/a_{\rm LNO}$ (full squares) with $c_{\rm LNO}=2c_{\rm SL}-c_{\rm RXO}$ and $a_{\rm LNO}=a_{\rm SL}$ and vs.\ the size of the X cation $r_X$ (Ref.~\onlinecite{Shannon1976}; open stars) for superlattices under tensile strain. The $c$-axis lattice parameter of \textit{R}XO was obtained from the Poisson ratio: $c_{\rm RXO}=\frac{2\nu}{\nu-1}(a_{\rm SL}-a_{\rm RXO}^{\rm bulk})+ a_{\rm RXO}^{\rm bulk}$ using $\nu=0.26$.\cite{Luo2088}}
\label{Fig:P_vs_a}
\end{figure}

Figure~\ref{Fig:P_vs_a} provides a synopsis of the orbital polarizations $P_{\rm av}$, $P_A$, and $P_B$ as a function of $a_{\rm SL}$. Note that the hole ratio $X_{\rm av}$ plotted in Fig.~\ref{Fig:LDIC_all}g is linearly related to $P_{\rm av}$ over the range investigated here, and within the given error bars. The polarization $P_A$ of the inner layers, which is less strongly affected by interfacial effects, depends linearly on $a_{\rm SL}$ over the entire measured range, including both the samples under tensile strain and the compressively strained LNO-LAO superlattice on LSAO, where $P_A$ is negative corresponding to an enhanced occupation of the $3z^2-r^2$ orbital. The fitted straight line crosses zero around $a_{\rm SL}=3.79$\,\AA, slightly below the pseudo-cubic bulk lattice constant of LNO of 3.838\,\AA.\cite{Munoz1992} We attribute this shift to the effect of confinement, yielding a slightly preferred $x^2-y^2$ occupation even for the inner layers (i.e.\ a small positive value of $P_A$). The strain dependence of both the energy splitting $\Delta e_g$ extracted from the cluster model discussed above and $P_A$ determined by orbital reflectometry thus indicate a simple linear orbital-lattice coupling.

Whereas strain alone induces orbital polarizations of up to $P_A \sim 25 \%$, the additional enhancement of the polarization in the outer LNO layers generated by interfacial effects falls into a comparatively narrow band of width $\sim 5 \%$ (grey area in Fig.~\ref{Fig:P_vs_a}). According to the DFT predictions,\cite{Hansmann2009,Hansmann2010,Han2010,Xiaoping2011} the orbital polarization at the LNO-\textit{R}XO interface is strongly influenced by the dimensional confinement of the conduction electrons and by the chemical composition of the blocking layers. The effect of confinement is expected to be similar in all of our superlattices, because the blocking layers have identical thicknesses (4 u.c.) and similar band-gaps. The effect of chemical composition is due to the hybridization between the Ni $3z^2-r^2$ and the O $p_z$-orbital of the apical oxygen at the interface, which in turn depends on the hybridization between the $s$-symmetry orbital of the X-ion with the O-$p_z$ states. The hybridization parameters are difficult to determine experimentally, but Han {\it et al.}\cite{Han2010} pointed out a close relationship between these quantities and parameters characterizing the lattice structure, including especially the O-X bond length, which is controlled by the size of the X cation, $r_{\rm X}$. Specifically, for large $r_{\rm X}$ (large O-X distances) the X-$s$/O-$p_z$ hybridization is expected to be reduced, and the Ni-$d_{3z^2-r^2}/$O-$p_z$ hybridization correspondingly enhanced, resulting in a larger enhancement of the orbital polarization at the interface, and vice versa. We have therefore plotted the ratio $P_B/P_A$ (which is a measure of the modulation of orbital polarization within the LNO layer stack) as a function of the lattice parameter ratio $c_{\rm LNO}/a_{\rm LNO}$ (inset of Fig.~\ref{Fig:P_vs_a}), which is approximately proportional to $r_{\rm X}$ (top axis in the inset of Fig.~\ref{Fig:P_vs_a}). The resulting trend of reduced interfacial enhancement of $P$ for smaller $r_{\rm X}$ is opposite to the trend predicted by the DFT calculations.\cite{Han2010,Xiaoping2011} A full crystallographic determination of the Ni-O and O-X distances and the Ni-O-X bond angle as well as corresponding DFT calculations are required to elucidate the origin of this discrepancy.

\section{Conclusion}

In summary, we have shown that tensile epitaxial strain can enhance the occupation of the $x^2-y^2$ orbital to $25\%$ in nickelate superlattices. The combined analysis of XAS and resonant reflectivity at the Ni $L$ edge revealed that strain induced by the lattice mismatch with the substrate has the largest effect on the orbital polarization. Especially when combined with other control parameters such as the conduction electron density in the LNO layers, the prospects for orbital engineering of the electronic properties of the nickelates and other oxide superlattices are therefore brighter than suggested by recent experimental and theoretical work.

\section*{Acknowledgements}

We thank G.~Khaliullin, O.~K.~Andersen, X.~Yang, O.~Peil, A.~Millis, H.~Boschker, and G.~A.~Sawatzky for fruitful discussions. We acknowledge the provision of synchrotron radiation and the assistance from W.~Mahler and B.~Zada at the Helmholtz-Zentrum Berlin-BESSY II as well as the financial support by the DFG via TRR80, project C1.

\appendix
\section*{Appendix: Growth and Characterization}\label{App:Growth and Characterization}

The superlattices were grown by pulsed-laser deposition from stoichiometric targets of LaNiO$_3$ (LNO) and \textit{R}XO$_3$ (\textit{R}XO) with \textit{R}\,=\,La, Dy, Gd and X\,=\,Al, Ga, and Sc, using a KrF excimer laser with 2\,Hz pulse rate and 1.6\,J/cm$^2$ energy density. All materials were deposited in 0.5\,mbar oxygen atmosphere at 730$^{\circ}$C and subsequently annealed in 1\,bar oxygen atmosphere at 690$^{\circ}$C for 30\,min, which we found to be the optimized conditions for the growth of LNO. Here we focus on superlattices with (4\,u.c.//4\,u.c.)$\times 8$ structure, where 1\,u.c. corresponds to one (pseudo)-cubic unit cell of LNO and \textit{R}XO, respectively.

\begin{figure}[t]
\includegraphics[width=0.8\columnwidth]{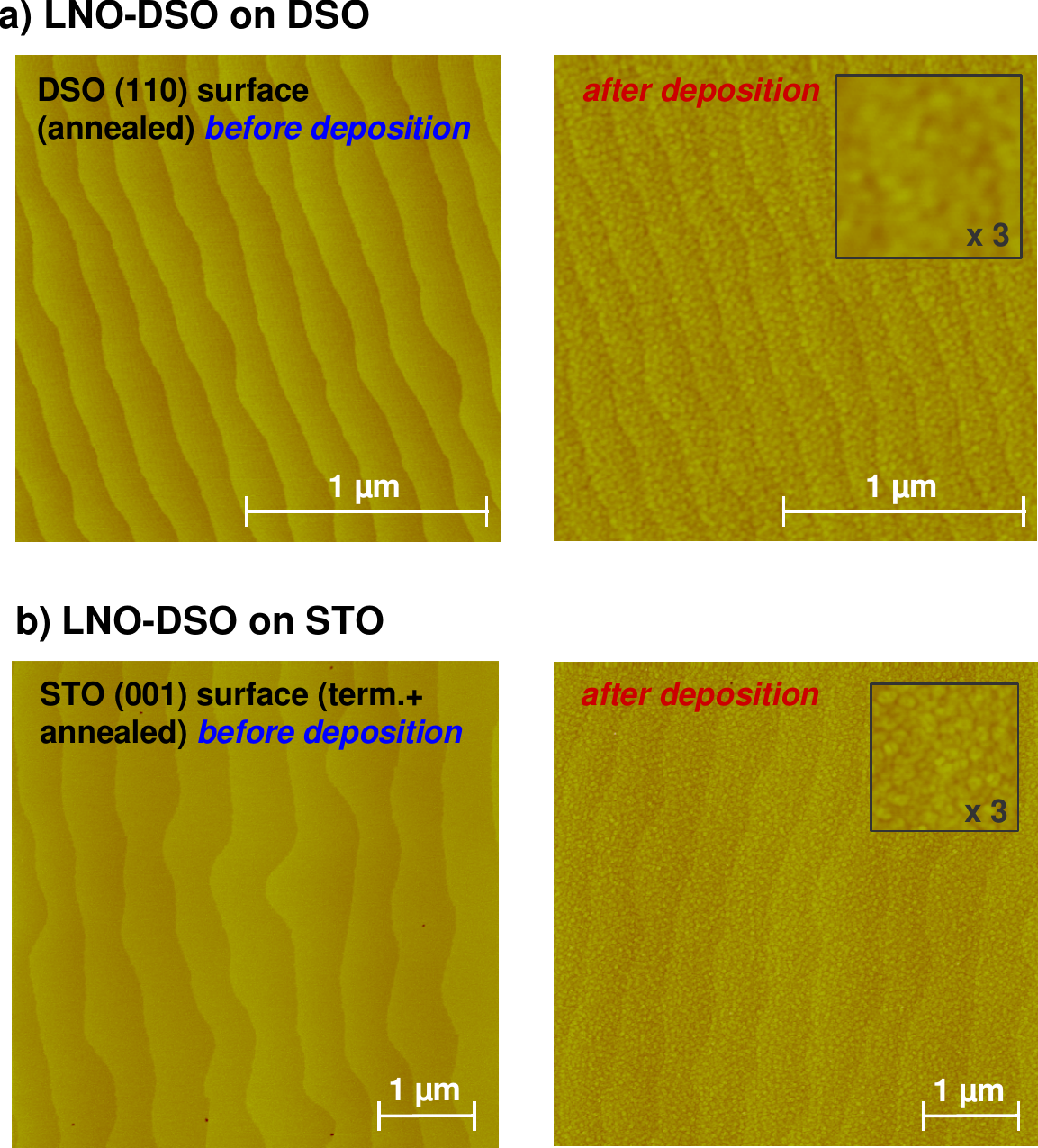}
\caption{\textit{Ex-situ} atomic force microscopy pictures of the specimen surfaces before (left) and after (right) deposition for a) LNO-DSO on DSO and b) LNO-DSO on STO.}
\label{Fig:AFM}
\end{figure}

\begin{figure*}[t]
\includegraphics[width=0.99\textwidth]{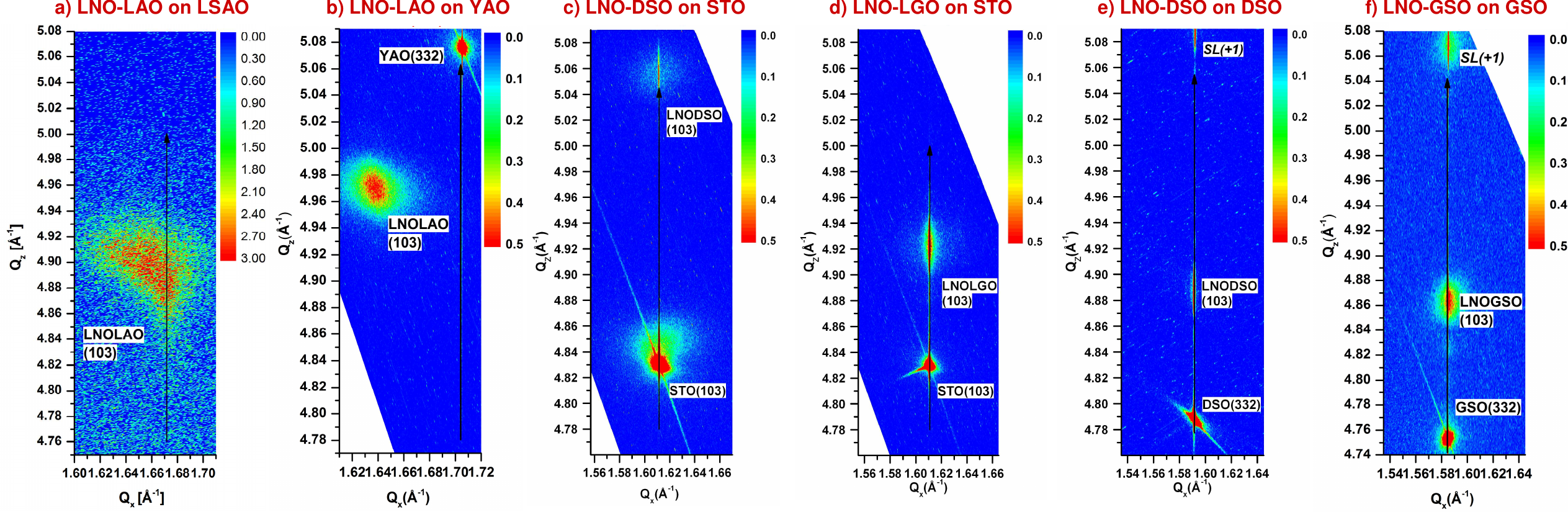}
\caption{Reciprocal space maps around the pseudo-cubic (103) peak position for (4//4)\,u.c.-SLs with composition (a) LNO-LAO on LSAO, (b) LNO-LAO on YAO, (c) LNO-DSO on STO, (d) LNO-LGO on STO, (e) LNO-DSO on DSO, and (f) LNO-GSO on GSO. The measurements were done using Cu $K_\alpha$ radiation ($E=8048$\,eV).}
\label{Fig:RSMaps}
\end{figure*}

Superlattices were deposited on single-crystalline substrates of [001]-oriented SrTiO$_3$ (STO), [001]-oriented LaSrAlO$_4$ (LSAO), and [110]-oriented YAO, DSO and GSO (see Table~\ref{Tab:Superlattices}). We used selected high-quality substrates of (5\,$\times$\,5\,$\times$\,0.5)\,mm size with very low mosaicity and miscut angles $<$\,0.3$^{\circ}$. Prior to deposition, the STO substrates were treated by a wet-chemical etching procedure with post-annealing in oxygen atmosphere at 900\,$^{\circ}$C (1\,h) to obtain TiO$_2$ termination and rearrangement of the surface.\cite{Kareev2008} The DSO substrates were annealed in oxygen atmosphere (1000$^{\circ}$C, 1h) to obtain surface rearrangement, however, the surface termination is expected to be a mixture of DyO and ScO$_2$ atomic layers.\cite{Kleibeuker2010} Atomic force microscopy pictures taken after these procedures revealed $\sim 4$\,\AA, ($\sim 1$\,u.c.) deep terraces with a lateral width of the order of 150\,nm and 500\,nm for DSO and STO substrates, respectively. Two representative pictures are shown in Fig.~\ref{Fig:AFM}. This surface morphology provides an optimized surface for the growth of superlattices with atomically flat interfaces. For all samples the surface morphology of the substrate is preserved after the deposition of the superlattice. The deposited LNO-\textit{R}XO surfaces show islands that indicate layer-by-layer epitaxial growth. The island diameters are approximately 30\,nm and 50\,nm for the LNO-DSO on DSO and LNO-DSO and STO superlattice, respectively (see insets in Fig.~\ref{Fig:AFM}). No treatment was applied to the LSAO and YAO substrates.

\begin{table}[t]
\begin{tabular}{l c c l l}
\hline\hline
\footnotesize{\textit{R}XO}                 & \footnotesize{Substrate}     & \footnotesize{SL structure}                                           & \footnotesize{$a_{\rm SL}$}  & \footnotesize{$c_{\rm SL}$} \\
              &               &   \footnotesize{(u.c.)}                                               &   \footnotesize{(\AA)}       & \footnotesize{(\AA)}        \\
\hline
\footnotesize{LAO} & \footnotesize{LSAO$_{0.5}$}       & \footnotesize{4.4$_{0.7}$[(4.1$_{0.6}$/4.2$_{0.7}$)x9]4.7$_{0.8}$}    & \footnotesize{3.750(5)}      & \footnotesize{3.840(2)} \\
\footnotesize{LAO} & \footnotesize{YAO}                & \footnotesize{(4//4)x8}                                               & \footnotesize{3.834(4)}      & \footnotesize{3.788(3)} \\
\footnotesize{DSO}  & \footnotesize{STO$_{0.8}$}        & \footnotesize{5.0$_{0.8}$[(3.9$_{1.3}$/4.0$_{1.7}$)x7]3.4$_{1.3}$}   & \footnotesize{3.905(2)}      & \footnotesize{3.870(8)} \\
\footnotesize{LGO}  & \footnotesize{STO$_{0.6}$}        & \footnotesize{4.0$_{0.5}$[(4.0$_{0.9}$/4.2$_{0.8}$)x7]5.1$_{1.6}$}   & \footnotesize{3.902(2)}      & \footnotesize{3.829(1)} \\
\footnotesize{DSO}  & \footnotesize{DSO$_{0.6}$}        & \footnotesize{4.7$_{1.4}$[(4.2$_{1.4}$/3.9$_{0.7}$)x7]3.6$_{1.3}$}   & \footnotesize{3.949(2)}      & \footnotesize{3.870(2)} \\
\footnotesize{GSO}  & \footnotesize{GSO$_{0.8}$}        & \footnotesize{3.7$_{0.6}$[(4.0$_{0.6}$/3.9$_{0.8}$)x7]4.4$_{1.2}$}   & \footnotesize{3.964(4)}      & \footnotesize{3.881(3)} \\
\hline\hline
\end{tabular}
\caption{Superlattice composition and structure, substrate material, and lattice parameter $a_{\rm SL}$ and $c_{\rm SL}$ of the investigated nickelate heterostructures (for details see text). The superlattice structure is obtained from fits to the non-resonant reflectivity (see text) and given in the following nomenclature: $d^{\rm LNObottom}_{\sigma}$[($d^{\rm LNO}_{\sigma}$/$d^{\rm RXO}_{\sigma}$)xM]$d^{\rm RXOtop}_{\sigma}$ with thickness $d$ and roughness $\sigma$ in u.c.\ calculated by dividing with $c_{\rm SL}$ and $M$ is number of repetitions of the bilayer. The roughness of the substrate is given in the corresponding column as an index, i.e.\ \textit{R}XO$_{\sigma}$. For the LNO-LAO on YAO no reflectivity measurements were performed.}
\label{Tab:Superlattices}
\end{table}

Thickness control of the individual superlattice layers was accomplished by counting laser pulses, using growth rates per pulse for LNO and \textit{R}XO obtained from previous tests with feedback from x-ray diffraction. Reciprocal space maps around the (103) pseudo-cubic Bragg peak positions are shown in Fig.~\ref{Fig:RSMaps}. From those measurements we obtained the in- and out-of-plane lattice constants $a_{\rm SL}$ and $c_{\rm SL}$, respectively, summarized in Table~\ref{Tab:Superlattices}. Note that these values are averaged values for both components of the superlattice, i.e.\ for LNO and \textit{R}XO layers, assuming a tetragonal crystal structure of the overlayer. The in-plane lattice constants of all superlattices fully match those of their substrates (vertical black arrows in Fig.~\ref{Fig:RSMaps}), except for the LNO-LAO on YAO superlattice which is partially relaxed.


\begin{thebibliography}{32}%
\makeatletter
\providecommand \@ifxundefined [1]{%
 \@ifx{#1\undefined}
}%
\providecommand \@ifnum [1]{%
 \ifnum #1\expandafter \@firstoftwo
 \else \expandafter \@secondoftwo
 \fi
}%
\providecommand \@ifx [1]{%
 \ifx #1\expandafter \@firstoftwo
 \else \expandafter \@secondoftwo
 \fi
}%
\providecommand \natexlab [1]{#1}%
\providecommand \enquote  [1]{``#1''}%
\providecommand \bibnamefont  [1]{#1}%
\providecommand \bibfnamefont [1]{#1}%
\providecommand \citenamefont [1]{#1}%
\providecommand \href@noop [0]{\@secondoftwo}%
\providecommand \href[0]{\begingroup \@sanitize@url \@href}%
\providecommand \@href[1]{\@@startlink{#1}\@@href}%
\providecommand \@@href[1]{\endgroup#1\@@endlink}%
\providecommand \@sanitize@url [0]{\catcode `\\12\catcode `\$12\catcode
  `\&12\catcode `\#12\catcode `\^12\catcode `\_12\catcode `\%12\relax}%
\providecommand \@@startlink[1]{}%
\providecommand \@@endlink[0]{}%
\providecommand \url  [0]{\begingroup\@sanitize@url \@url }%
\providecommand \@url [1]{\endgroup\@href {#1}{\urlprefix }}%
\providecommand \urlprefix  [0]{URL }%
\providecommand \Eprint [0]{\href}%
\providecommand \doibase [0]{http://dx.doi.org}%
\providecommand \selectlanguage [0]{\@gobble}%
\providecommand \bibinfo  [0]{\@secondoftwo}%
\providecommand \bibfield  [0]{\@secondoftwo}%
\providecommand \translation [1]{[#1]}%
\providecommand \BibitemOpen [0]{}%
\providecommand \bibitemStop [0]{}%
\providecommand \bibitemNoStop [0]{.\EOS\space}%
\providecommand \EOS [0]{\spacefactor3000\relax}%
\providecommand \BibitemShut  [1]{\csname bibitem#1\endcsname}%
\let\auto@bib@innerbib\@empty
\bibitem [{\citenamefont {Chaloupka}\ and\ \citenamefont
  {Khaliullin}(2008)}]{Chaloupka2008}%
  \BibitemOpen
  \bibfield  {author} {\bibinfo {author} {\bibfnamefont {J.}~\bibnamefont
  {Chaloupka}}\ and\ \bibinfo {author} {\bibfnamefont {G.}~\bibnamefont
  {Khaliullin}},\
  }\href{http://link.aps.org/doi/10.1103/PhysRevLett.100.016404} {\bibfield
  {journal} {\bibinfo  {journal} {Phys. Rev. Lett.}\ }\textbf {\bibinfo
  {volume} {100}},\ \bibinfo {pages} {016404} (\bibinfo {year}
  {2008})}\BibitemShut {NoStop}%
\bibitem [{\citenamefont {Hansmann}\ \emph {\textit{et~al.}}(2009)\citenamefont
  {Hansmann}, \citenamefont {Yang}, \citenamefont {Toschi}, \citenamefont
  {Khaliullin}, \citenamefont {Andersen},\ and\ \citenamefont
  {Held}}]{Hansmann2009}%
  \BibitemOpen
  \bibfield  {author} {\bibinfo {author} {\bibfnamefont {P.}~\bibnamefont
  {Hansmann}}, \bibinfo {author} {\bibfnamefont {X.}~\bibnamefont {Yang}},
  \bibinfo {author} {\bibfnamefont {A.}~\bibnamefont {Toschi}}, \bibinfo
  {author} {\bibfnamefont {G.}~\bibnamefont {Khaliullin}}, \bibinfo {author}
  {\bibfnamefont {O.~K.}\ \bibnamefont {Andersen}}, \ and\ \bibinfo {author}
  {\bibfnamefont {K.}~\bibnamefont {Held}},\
  }\href{http://link.aps.org/doi/10.1103/PhysRevLett.103.016401} {\bibfield
  {journal} {\bibinfo  {journal} {Phys. Rev. Lett.}\ }\textbf {\bibinfo
  {volume} {103}},\ \bibinfo {pages} {016401} (\bibinfo {year}
  {2009})}\BibitemShut {NoStop}%
\bibitem [{\citenamefont {Han}\ \emph {\textit{et~al.}}(2010)\citenamefont
  {Han}, \citenamefont {Marianetti},\ and\ \citenamefont {Millis}}]{Han2010}%
  \BibitemOpen
  \bibfield  {author} {\bibinfo {author} {\bibfnamefont {M.~J.}\ \bibnamefont
  {Han}}, \bibinfo {author} {\bibfnamefont {C.~A.}\ \bibnamefont {Marianetti}},
  \ and\ \bibinfo {author} {\bibfnamefont {A.~J.}\ \bibnamefont {Millis}},\
  }\href{http://link.aps.org/doi/10.1103/PhysRevB.82.134408} {\bibfield
  {journal} {\bibinfo  {journal} {Phys. Rev. B}\ }\textbf {\bibinfo {volume}
  {82}},\ \bibinfo {pages} {134408} (\bibinfo {year} {2010})}\BibitemShut
  {NoStop}%
\bibitem [{\citenamefont {Hansmann}\ \emph {\textit{et~al.}}(2010)\citenamefont
  {Hansmann}, \citenamefont {Toschi}, \citenamefont {Yang}, \citenamefont
  {Andersen},\ and\ \citenamefont {Held}}]{Hansmann2010}%
  \BibitemOpen
  \bibfield  {author} {\bibinfo {author} {\bibfnamefont {P.}~\bibnamefont
  {Hansmann}}, \bibinfo {author} {\bibfnamefont {A.}~\bibnamefont {Toschi}},
  \bibinfo {author} {\bibfnamefont {X.}~\bibnamefont {Yang}}, \bibinfo {author}
  {\bibfnamefont {O.~K.}\ \bibnamefont {Andersen}}, \ and\ \bibinfo {author}
  {\bibfnamefont {K.}~\bibnamefont {Held}},\
  }\href{http://link.aps.org/doi/10.1103/PhysRevB.82.235123} {\bibfield
  {journal} {\bibinfo  {journal} {Phys. Rev. B}\ }\textbf {\bibinfo {volume}
  {82}},\ \bibinfo {pages} {235123} (\bibinfo {year} {2010})}\BibitemShut
  {NoStop}%
\bibitem [{\citenamefont {Yang}\ \emph {\textit{et~al.}}()\citenamefont {Yang},
  \citenamefont {Hansmann}, \citenamefont {Toschi}, \citenamefont {Held},
  \citenamefont {Khaliullin},\ and\ \citenamefont {Andersen}}]{Xiaoping2011}%
  \BibitemOpen
  \bibfield  {author} {\bibinfo {author} {\bibfnamefont {X.}~\bibnamefont
  {Yang}}, \bibinfo {author} {\bibfnamefont {P.}~\bibnamefont {Hansmann}},
  \bibinfo {author} {\bibfnamefont {A.}~\bibnamefont {Toschi}}, \bibinfo
  {author} {\bibfnamefont {K.}~\bibnamefont {Held}}, \bibinfo {author}
  {\bibfnamefont {G.}~\bibnamefont {Khaliullin}}, \ and\ \bibinfo {author}
  {\bibfnamefont {O.~K.}\ \bibnamefont {Andersen}},\ }\href@noop {} {}\bibinfo
  {note} {unpublished; see also O.~K.~Andersen, APS March Meeting, (Pittsburgh,
  Pennsylvania, 2009)}\BibitemShut {NoStop}%
\bibitem [{\citenamefont {Blanca-Romero}\ and\ \citenamefont
  {Pentcheva}(2011)}]{Blanca2011}%
  \BibitemOpen
  \bibfield  {author} {\bibinfo {author} {\bibfnamefont {A.}~\bibnamefont
  {Blanca-Romero}}\ and\ \bibinfo {author} {\bibfnamefont {R.}~\bibnamefont
  {Pentcheva}},\ }\href{http://link.aps.org/doi/10.1103/PhysRevB.84.195450}
  {\bibfield  {journal} {\bibinfo  {journal} {Phys. Rev. B}\ }\textbf {\bibinfo
  {volume} {84}},\ \bibinfo {pages} {195450} (\bibinfo {year}
  {2011})}\BibitemShut {NoStop}%
\bibitem [{\citenamefont {Han}\ \emph {\textit{et~al.}}(2011)\citenamefont
  {Han}, \citenamefont {Wang}, \citenamefont {Marianetti},\ and\ \citenamefont
  {Millis}}]{Han2011}%
  \BibitemOpen
  \bibfield  {author} {\bibinfo {author} {\bibfnamefont {M.~J.}\ \bibnamefont
  {Han}}, \bibinfo {author} {\bibfnamefont {X.}~\bibnamefont {Wang}}, \bibinfo
  {author} {\bibfnamefont {C.~A.}\ \bibnamefont {Marianetti}}, \ and\ \bibinfo
  {author} {\bibfnamefont {A.~J.}\ \bibnamefont {Millis}},\
  }\href{http://link.aps.org/doi/10.1103/PhysRevLett.107.206804} {\bibfield
  {journal} {\bibinfo  {journal} {Phys. Rev. Lett.}\ }\textbf {\bibinfo
  {volume} {107}},\ \bibinfo {pages} {206804} (\bibinfo {year}
  {2011})}\BibitemShut {NoStop}%
\bibitem [{\citenamefont {Han}\ and\ \citenamefont {van
  Veenendaal}(2012)}]{Han2012}%
  \BibitemOpen
  \bibfield  {author} {\bibinfo {author} {\bibfnamefont {M.~J.}\ \bibnamefont
  {Han}}\ and\ \bibinfo {author} {\bibfnamefont {M.}~\bibnamefont {van
  Veenendaal}},\ }\href{http://link.aps.org/doi/10.1103/PhysRevB.85.195102}
  {\bibfield  {journal} {\bibinfo  {journal} {Phys. Rev. B}\ }\textbf {\bibinfo
  {volume} {85}},\ \bibinfo {pages} {195102} (\bibinfo {year}
  {2012})}\BibitemShut {NoStop}%
\bibitem [{\citenamefont {Chakhalian}\ \emph
  {\textit{et~al.}}(2011)\citenamefont {Chakhalian}, \citenamefont
  {Rondinelli}, \citenamefont {Liu}, \citenamefont {Gray}, \citenamefont
  {Kareev}, \citenamefont {Moon}, \citenamefont {Prasai}, \citenamefont {Cohn},
  \citenamefont {Varela}, \citenamefont {Tung}, \citenamefont {Bedzyk},
  \citenamefont {Altendorf}, \citenamefont {Strigari}, \citenamefont
  {Dabrowski}, \citenamefont {Tjeng}, \citenamefont {Ryan},\ and\ \citenamefont
  {Freeland}}]{Chakhalian2011}%
  \BibitemOpen
  \bibfield  {author} {\bibinfo {author} {\bibfnamefont {J.}~\bibnamefont
  {Chakhalian}}, \bibinfo {author} {\bibfnamefont {J.~M.}\ \bibnamefont
  {Rondinelli}}, \bibinfo {author} {\bibfnamefont {J.}~\bibnamefont {Liu}},
  \bibinfo {author} {\bibfnamefont {B.~A.}\ \bibnamefont {Gray}}, \bibinfo
  {author} {\bibfnamefont {M.}~\bibnamefont {Kareev}}, \bibinfo {author}
  {\bibfnamefont {E.~J.}\ \bibnamefont {Moon}}, \bibinfo {author}
  {\bibfnamefont {N.}~\bibnamefont {Prasai}}, \bibinfo {author} {\bibfnamefont
  {J.~L.}\ \bibnamefont {Cohn}}, \bibinfo {author} {\bibfnamefont
  {M.}~\bibnamefont {Varela}}, \bibinfo {author} {\bibfnamefont {I.~C.}\
  \bibnamefont {Tung}}, \bibinfo {author} {\bibfnamefont {M.~J.}\ \bibnamefont
  {Bedzyk}}, \bibinfo {author} {\bibfnamefont {S.~G.}\ \bibnamefont
  {Altendorf}}, \bibinfo {author} {\bibfnamefont {F.}~\bibnamefont {Strigari}},
  \bibinfo {author} {\bibfnamefont {B.}~\bibnamefont {Dabrowski}}, \bibinfo
  {author} {\bibfnamefont {L.~H.}\ \bibnamefont {Tjeng}}, \bibinfo {author}
  {\bibfnamefont {P.~J.}\ \bibnamefont {Ryan}}, \ and\ \bibinfo {author}
  {\bibfnamefont {J.~W.}\ \bibnamefont {Freeland}},\
  }\href{http://link.aps.org/doi/10.1103/PhysRevLett.107.116805} {\bibfield
  {journal} {\bibinfo  {journal} {Phys. Rev. Lett.}\ }\textbf {\bibinfo
  {volume} {107}},\ \bibinfo {pages} {116805} (\bibinfo {year}
  {2011})}\BibitemShut {NoStop}%
\bibitem [{\citenamefont {Freeland}\ \emph {\textit{et~al.}}(2011)\citenamefont
  {Freeland}, \citenamefont {Liu}, \citenamefont {Kareev}, \citenamefont
  {Gray}, \citenamefont {Kim}, \citenamefont {Ryan}, \citenamefont
  {Pentcheva},\ and\ \citenamefont {Chakhalian}}]{Freeland2011}%
  \BibitemOpen
  \bibfield  {author} {\bibinfo {author} {\bibfnamefont {J.~W.}\ \bibnamefont
  {Freeland}}, \bibinfo {author} {\bibfnamefont {J.}~\bibnamefont {Liu}},
  \bibinfo {author} {\bibfnamefont {M.}~\bibnamefont {Kareev}}, \bibinfo
  {author} {\bibfnamefont {B.}~\bibnamefont {Gray}}, \bibinfo {author}
  {\bibfnamefont {J.~W.}\ \bibnamefont {Kim}}, \bibinfo {author} {\bibfnamefont
  {P.}~\bibnamefont {Ryan}}, \bibinfo {author} {\bibfnamefont {R.}~\bibnamefont
  {Pentcheva}}, \ and\ \bibinfo {author} {\bibfnamefont {J.}~\bibnamefont
  {Chakhalian}},\ }\href{http://stacks.iop.org/0295-5075/96/i=5/a=57004}
  {\bibfield  {journal} {\bibinfo  {journal} {EPL (Europhysics Letters)}\
  }\textbf {\bibinfo {volume} {96}},\ \bibinfo {pages} {57004} (\bibinfo {year}
  {2011})}\BibitemShut {NoStop}%
\bibitem [{\citenamefont {Benckiser}\ \emph
  {\textit{et~al.}}(2011)\citenamefont {Benckiser}, \citenamefont {Haverkort},
  \citenamefont {Brueck}, \citenamefont {Goering}, \citenamefont {Macke},
  \citenamefont {Frano}, \citenamefont {Yang}, \citenamefont {Andersen},
  \citenamefont {Cristiani}, \citenamefont {Habermeier}, \citenamefont {Boris},
  \citenamefont {Zegkinoglou}, \citenamefont {Wochner}, \citenamefont {Kim},
  \citenamefont {Hinkov},\ and\ \citenamefont {Keimer}}]{Benckiser2011}%
  \BibitemOpen
  \bibfield  {author} {\bibinfo {author} {\bibfnamefont {E.}~\bibnamefont
  {Benckiser}}, \bibinfo {author} {\bibfnamefont {M.~W.}\ \bibnamefont
  {Haverkort}}, \bibinfo {author} {\bibfnamefont {S.}~\bibnamefont {Brueck}},
  \bibinfo {author} {\bibfnamefont {E.}~\bibnamefont {Goering}}, \bibinfo
  {author} {\bibfnamefont {S.}~\bibnamefont {Macke}}, \bibinfo {author}
  {\bibfnamefont {A.}~\bibnamefont {Frano}}, \bibinfo {author} {\bibfnamefont
  {X.}~\bibnamefont {Yang}}, \bibinfo {author} {\bibfnamefont {O.~K.}\
  \bibnamefont {Andersen}}, \bibinfo {author} {\bibfnamefont {G.}~\bibnamefont
  {Cristiani}}, \bibinfo {author} {\bibfnamefont {H.-U.}\ \bibnamefont
  {Habermeier}}, \bibinfo {author} {\bibfnamefont {A.~V.}\ \bibnamefont
  {Boris}}, \bibinfo {author} {\bibfnamefont {I.}~\bibnamefont {Zegkinoglou}},
  \bibinfo {author} {\bibfnamefont {P.}~\bibnamefont {Wochner}}, \bibinfo
  {author} {\bibfnamefont {H.-J.}\ \bibnamefont {Kim}}, \bibinfo {author}
  {\bibfnamefont {V.}~\bibnamefont {Hinkov}}, \ and\ \bibinfo {author}
  {\bibfnamefont {B.}~\bibnamefont {Keimer}},\
  }\href{\doibase/10.1038/NMAT2958} {\bibfield  {journal} {\bibinfo  {journal}
  {Nature Materials}\ }\textbf {\bibinfo {volume} {10}},\ \bibinfo {pages}
  {189} (\bibinfo {year} {2011})}\BibitemShut {NoStop}%
\bibitem [{\citenamefont {Detemple}\ \emph {\textit{et~al.}}(2011)\citenamefont
  {Detemple}, \citenamefont {Ramasse}, \citenamefont {Sigle}, \citenamefont
  {Cristiani}, \citenamefont {Habermeier}, \citenamefont {Benckiser},
  \citenamefont {Boris}, \citenamefont {Frano}, \citenamefont {Wochner},
  \citenamefont {Wu}, \citenamefont {Keimer},\ and\ \citenamefont {van
  Aken}}]{Detemple2011}%
  \BibitemOpen
  \bibfield  {author} {\bibinfo {author} {\bibfnamefont {E.}~\bibnamefont
  {Detemple}}, \bibinfo {author} {\bibfnamefont {Q.~M.}\ \bibnamefont
  {Ramasse}}, \bibinfo {author} {\bibfnamefont {W.}~\bibnamefont {Sigle}},
  \bibinfo {author} {\bibfnamefont {G.}~\bibnamefont {Cristiani}}, \bibinfo
  {author} {\bibfnamefont {H.-U.}\ \bibnamefont {Habermeier}}, \bibinfo
  {author} {\bibfnamefont {E.}~\bibnamefont {Benckiser}}, \bibinfo {author}
  {\bibfnamefont {A.~V.}\ \bibnamefont {Boris}}, \bibinfo {author}
  {\bibfnamefont {A.}~\bibnamefont {Frano}}, \bibinfo {author} {\bibfnamefont
  {P.}~\bibnamefont {Wochner}}, \bibinfo {author} {\bibfnamefont
  {M.}~\bibnamefont {Wu}}, \bibinfo {author} {\bibfnamefont {B.}~\bibnamefont
  {Keimer}}, \ and\ \bibinfo {author} {\bibfnamefont {P.~A.}\ \bibnamefont {van
  Aken}},\ }\href{http://link.aip.org/link/?APL/99/211903/1} {\bibfield
  {journal} {\bibinfo  {journal} {Applied Physics Letters}\ }\textbf {\bibinfo
  {volume} {99}},\ \bibinfo {eid} {211903} (\bibinfo {year}
  {2011})}\BibitemShut {NoStop}%
\bibitem [{\citenamefont {Detemple}\ \emph {\textit{et~al.}}(2012)\citenamefont
  {Detemple}, \citenamefont {Ramasse}, \citenamefont {Sigle}, \citenamefont
  {Cristiani}, \citenamefont {Habermeier}, \citenamefont {Keimer},\ and\
  \citenamefont {van Aken}}]{Detemple2012}%
  \BibitemOpen
  \bibfield  {author} {\bibinfo {author} {\bibfnamefont {E.}~\bibnamefont
  {Detemple}}, \bibinfo {author} {\bibfnamefont {Q.~M.}\ \bibnamefont
  {Ramasse}}, \bibinfo {author} {\bibfnamefont {W.}~\bibnamefont {Sigle}},
  \bibinfo {author} {\bibfnamefont {G.}~\bibnamefont {Cristiani}}, \bibinfo
  {author} {\bibfnamefont {H.-U.}\ \bibnamefont {Habermeier}}, \bibinfo
  {author} {\bibfnamefont {B.}~\bibnamefont {Keimer}}, \ and\ \bibinfo {author}
  {\bibfnamefont {P.~A.}\ \bibnamefont {van Aken}},\
  }\href{http://link.aip.org/link/?JAP/112/013509/1} {\bibfield  {journal}
  {\bibinfo  {journal} {Journal of Applied Physics}\ }\textbf {\bibinfo
  {volume} {112}},\ \bibinfo {eid} {013509} (\bibinfo {year}
  {2012})}\BibitemShut {NoStop}%
\bibitem [{\citenamefont {Detemple}(2013)}]{DetempleThesis}%
  \BibitemOpen
  \bibfield  {author} {\bibinfo {author} {\bibfnamefont {E.}~\bibnamefont
  {Detemple}},\ }\emph {\bibinfo {title} {Charakterisierung von
  LaNiO$_3$-basierten Übergittern mittels transmissionselektronmikroskopischer
  Methoden}},\ \href@noop {} {Ph.D. thesis},\ \bibinfo  {school} {Technische
  Universität Darmstadt} (\bibinfo {year} {2013})\BibitemShut {NoStop}%
\bibitem [{\citenamefont {Bruck}\ \emph {\textit{et~al.}}(2008)\citenamefont
  {Bruck}, \citenamefont {Bauknecht}, \citenamefont {Ludescher}, \citenamefont
  {Goering},\ and\ \citenamefont {Schutz}}]{Brueck2008}%
  \BibitemOpen
  \bibfield  {author} {\bibinfo {author} {\bibfnamefont {S.}~\bibnamefont
  {Bruck}}, \bibinfo {author} {\bibfnamefont {S.}~\bibnamefont {Bauknecht}},
  \bibinfo {author} {\bibfnamefont {B.}~\bibnamefont {Ludescher}}, \bibinfo
  {author} {\bibfnamefont {E.}~\bibnamefont {Goering}}, \ and\ \bibinfo
  {author} {\bibfnamefont {G.}~\bibnamefont {Schutz}},\
  }\href{http://link.aip.org/link/?RSI/79/083109/1} {\bibfield  {journal}
  {\bibinfo  {journal} {Rev.\ Sci.\ Instr.}\ }\textbf {\bibinfo {volume}
  {79}},\ \bibinfo {eid} {083109} (\bibinfo {year} {2008})}\BibitemShut
  {NoStop}%
\bibitem [{\citenamefont {Boris}\ \emph {\textit{et~al.}}(2011)\citenamefont
  {Boris}, \citenamefont {Matiks}, \citenamefont {Benckiser}, \citenamefont
  {Frano}, \citenamefont {Popovich}, \citenamefont {Hinkov}, \citenamefont
  {Wochner}, \citenamefont {Castro-Colin}, \citenamefont {Detemple},
  \citenamefont {Malik}, \citenamefont {Bernhard}, \citenamefont {Prokscha},
  \citenamefont {Suter}, \citenamefont {Salman}, \citenamefont {Morenzoni},
  \citenamefont {Cristiani}, \citenamefont {Habermeier},\ and\ \citenamefont
  {Keimer}}]{Boris2011}%
  \BibitemOpen
  \bibfield  {author} {\bibinfo {author} {\bibfnamefont {A.~V.}\ \bibnamefont
  {Boris}}, \bibinfo {author} {\bibfnamefont {Y.}~\bibnamefont {Matiks}},
  \bibinfo {author} {\bibfnamefont {E.}~\bibnamefont {Benckiser}}, \bibinfo
  {author} {\bibfnamefont {A.}~\bibnamefont {Frano}}, \bibinfo {author}
  {\bibfnamefont {P.}~\bibnamefont {Popovich}}, \bibinfo {author}
  {\bibfnamefont {V.}~\bibnamefont {Hinkov}}, \bibinfo {author} {\bibfnamefont
  {P.}~\bibnamefont {Wochner}}, \bibinfo {author} {\bibfnamefont
  {M.}~\bibnamefont {Castro-Colin}}, \bibinfo {author} {\bibfnamefont
  {E.}~\bibnamefont {Detemple}}, \bibinfo {author} {\bibfnamefont {V.~K.}\
  \bibnamefont {Malik}}, \bibinfo {author} {\bibfnamefont {C.}~\bibnamefont
  {Bernhard}}, \bibinfo {author} {\bibfnamefont {T.}~\bibnamefont {Prokscha}},
  \bibinfo {author} {\bibfnamefont {A.}~\bibnamefont {Suter}}, \bibinfo
  {author} {\bibfnamefont {Z.}~\bibnamefont {Salman}}, \bibinfo {author}
  {\bibfnamefont {E.}~\bibnamefont {Morenzoni}}, \bibinfo {author}
  {\bibfnamefont {G.}~\bibnamefont {Cristiani}}, \bibinfo {author}
  {\bibfnamefont {H.-U.}\ \bibnamefont {Habermeier}}, \ and\ \bibinfo {author}
  {\bibfnamefont {B.}~\bibnamefont {Keimer}},\
  }\href{http://www.sciencemag.org/content/332/6032/937.abstract} {\bibfield
  {journal} {\bibinfo  {journal} {Science}\ }\textbf {\bibinfo {volume}
  {332}},\ \bibinfo {pages} {937} (\bibinfo {year} {2011})}\BibitemShut
  {NoStop}%
\bibitem [{\citenamefont {Frano}\ \emph {\textit{et~al.}}(2013)\citenamefont
  {Frano}, \citenamefont {Schierle}, \citenamefont {Haverkort}, \citenamefont
  {Lu}, \citenamefont {Wu}, \citenamefont {Blanco-Canosa}, \citenamefont
  {Nwankwo}, \citenamefont {Boris}, \citenamefont {Wochner}, \citenamefont
  {Cristiani}, \citenamefont {Habermeier}, \citenamefont {Logvenov},
  \citenamefont {Hinkov}, \citenamefont {Benckiser}, \citenamefont {Weschke},\
  and\ \citenamefont {Keimer}}]{Frano2012}%
  \BibitemOpen
  \bibfield  {author} {\bibinfo {author} {\bibfnamefont {A.}~\bibnamefont
  {Frano}}, \bibinfo {author} {\bibfnamefont {E.}~\bibnamefont {Schierle}},
  \bibinfo {author} {\bibfnamefont {M.~W.}\ \bibnamefont {Haverkort}}, \bibinfo
  {author} {\bibfnamefont {Y.}~\bibnamefont {Lu}}, \bibinfo {author}
  {\bibfnamefont {M.}~\bibnamefont {Wu}}, \bibinfo {author} {\bibfnamefont
  {S.}~\bibnamefont {Blanco-Canosa}}, \bibinfo {author} {\bibfnamefont
  {U.}~\bibnamefont {Nwankwo}}, \bibinfo {author} {\bibfnamefont {A.~V.}\
  \bibnamefont {Boris}}, \bibinfo {author} {\bibfnamefont {P.}~\bibnamefont
  {Wochner}}, \bibinfo {author} {\bibfnamefont {G.}~\bibnamefont {Cristiani}},
  \bibinfo {author} {\bibfnamefont {H.~U.}\ \bibnamefont {Habermeier}},
  \bibinfo {author} {\bibfnamefont {G.}~\bibnamefont {Logvenov}}, \bibinfo
  {author} {\bibfnamefont {V.}~\bibnamefont {Hinkov}}, \bibinfo {author}
  {\bibfnamefont {E.}~\bibnamefont {Benckiser}}, \bibinfo {author}
  {\bibfnamefont {E.}~\bibnamefont {Weschke}}, \ and\ \bibinfo {author}
  {\bibfnamefont {B.}~\bibnamefont {Keimer}},\
  }\href{http://arxiv.org/abs/1304.1469} {\bibfield  {journal} {\bibinfo
  {journal} {arXiv:1304.1469}\ } (\bibinfo {year} {2013})}\BibitemShut
  {NoStop}%
\bibitem [{\citenamefont {van~der Laan}(1994)}]{vanderLaan1994}%
  \BibitemOpen
  \bibfield  {author} {\bibinfo {author} {\bibfnamefont {G.}~\bibnamefont
  {van~der Laan}},\ }\href{http://jpsj.ipap.jp/link?JPSJ/63/2393/} {\bibfield
  {journal} {\bibinfo  {journal} {J.\ Phys.\ Soc.\ Jpn.}\ }\textbf {\bibinfo
  {volume} {63}},\ \bibinfo {pages} {2393} (\bibinfo {year}
  {1994})}\BibitemShut {NoStop}%
\bibitem [{\citenamefont {de~Groot}\ \emph {\textit{et~al.}}(1990)\citenamefont
  {de~Groot}, \citenamefont {Fuggle}, \citenamefont {Thole},\ and\
  \citenamefont {Sawatzky}}]{deGroot1990}%
  \BibitemOpen
  \bibfield  {author} {\bibinfo {author} {\bibfnamefont {F.~M.~F.}\
  \bibnamefont {de~Groot}}, \bibinfo {author} {\bibfnamefont {J.~C.}\
  \bibnamefont {Fuggle}}, \bibinfo {author} {\bibfnamefont {B.~T.}\
  \bibnamefont {Thole}}, \ and\ \bibinfo {author} {\bibfnamefont {G.~A.}\
  \bibnamefont {Sawatzky}},\
  }\href{http://link.aps.org/doi/10.1103/PhysRevB.42.5459} {\bibfield
  {journal} {\bibinfo  {journal} {Phys. Rev. B}\ }\textbf {\bibinfo {volume}
  {42}},\ \bibinfo {pages} {5459} (\bibinfo {year} {1990})}\BibitemShut
  {NoStop}%
\bibitem [{\citenamefont {Tanaka}\ and\ \citenamefont {Jo}(1994)}]{Tanaka1994}%
  \BibitemOpen
  \bibfield  {author} {\bibinfo {author} {\bibfnamefont {A.}~\bibnamefont
  {Tanaka}}\ and\ \bibinfo {author} {\bibfnamefont {T.}~\bibnamefont {Jo}},\
  }\href{http://jpsj.ipap.jp/link?JPSJ/63/2788/} {\bibfield  {journal}
  {\bibinfo  {journal} {J. Phys. Soc. Jpn.}\ }\textbf {\bibinfo {volume}
  {63}},\ \bibinfo {pages} {2788} (\bibinfo {year} {1994})}\BibitemShut
  {NoStop}%
\bibitem [{\citenamefont {Sch\"u\ss{}ler-Langeheine}\ \emph
  {\textit{et~al.}}(2005)\citenamefont {Sch\"u\ss{}ler-Langeheine},
  \citenamefont {Schlappa}, \citenamefont {Tanaka}, \citenamefont {Hu},
  \citenamefont {Chang}, \citenamefont {Schierle}, \citenamefont {Benomar},
  \citenamefont {Ott}, \citenamefont {Weschke}, \citenamefont {Kaindl},
  \citenamefont {Friedt}, \citenamefont {Sawatzky}, \citenamefont {Lin},
  \citenamefont {Chen}, \citenamefont {Braden},\ and\ \citenamefont
  {Tjeng}}]{Schuessler2005}%
  \BibitemOpen
  \bibfield  {author} {\bibinfo {author} {\bibfnamefont {C.}~\bibnamefont
  {Sch\"u\ss{}ler-Langeheine}}, \bibinfo {author} {\bibfnamefont
  {J.}~\bibnamefont {Schlappa}}, \bibinfo {author} {\bibfnamefont
  {A.}~\bibnamefont {Tanaka}}, \bibinfo {author} {\bibfnamefont
  {Z.}~\bibnamefont {Hu}}, \bibinfo {author} {\bibfnamefont {C.~F.}\
  \bibnamefont {Chang}}, \bibinfo {author} {\bibfnamefont {E.}~\bibnamefont
  {Schierle}}, \bibinfo {author} {\bibfnamefont {M.}~\bibnamefont {Benomar}},
  \bibinfo {author} {\bibfnamefont {H.}~\bibnamefont {Ott}}, \bibinfo {author}
  {\bibfnamefont {E.}~\bibnamefont {Weschke}}, \bibinfo {author} {\bibfnamefont
  {G.}~\bibnamefont {Kaindl}}, \bibinfo {author} {\bibfnamefont
  {O.}~\bibnamefont {Friedt}}, \bibinfo {author} {\bibfnamefont {G.~A.}\
  \bibnamefont {Sawatzky}}, \bibinfo {author} {\bibfnamefont {H.-J.}\
  \bibnamefont {Lin}}, \bibinfo {author} {\bibfnamefont {C.~T.}\ \bibnamefont
  {Chen}}, \bibinfo {author} {\bibfnamefont {M.}~\bibnamefont {Braden}}, \ and\
  \bibinfo {author} {\bibfnamefont {L.~H.}\ \bibnamefont {Tjeng}},\
  }\href{http://link.aps.org/doi/10.1103/PhysRevLett.95.156402} {\bibfield
  {journal} {\bibinfo  {journal} {Phys. Rev. Lett.}\ }\textbf {\bibinfo
  {volume} {95}},\ \bibinfo {pages} {156402} (\bibinfo {year}
  {2005})}\BibitemShut {NoStop}%
\bibitem [{\citenamefont {Garc\`{\i}a-Mu$\tilde{\rm n}$oz}\ \emph
  {\textit{et~al.}}(1992)\citenamefont {Garc\`{\i}a-Mu$\tilde{\rm n}$oz},
  \citenamefont {Rodr\`{\i}guez-Carvajal}, \citenamefont {Lacorre},\ and\
  \citenamefont {Torrance}}]{Munoz1992}%
  \BibitemOpen
  \bibfield  {author} {\bibinfo {author} {\bibfnamefont {J.~L.}\ \bibnamefont
  {Garc\`{\i}a-Mu$\tilde{\rm n}$oz}}, \bibinfo {author} {\bibfnamefont
  {J.}~\bibnamefont {Rodr\`{\i}guez-Carvajal}}, \bibinfo {author}
  {\bibfnamefont {P.}~\bibnamefont {Lacorre}}, \ and\ \bibinfo {author}
  {\bibfnamefont {J.~B.}\ \bibnamefont {Torrance}},\
  }\href{http://link.aps.org/doi/10.1103/PhysRevB.46.4414} {\bibfield
  {journal} {\bibinfo  {journal} {Phys. Rev. B}\ }\textbf {\bibinfo {volume}
  {46}},\ \bibinfo {pages} {4414} (\bibinfo {year} {1992})}\BibitemShut
  {NoStop}%
\bibitem [{\citenamefont {Andersen}\ and\ \citenamefont
  {Saha-Dasgupta}(2000)}]{Andersen2000}%
  \BibitemOpen
  \bibfield  {author} {\bibinfo {author} {\bibfnamefont {O.~K.}\ \bibnamefont
  {Andersen}}\ and\ \bibinfo {author} {\bibfnamefont {T.}~\bibnamefont
  {Saha-Dasgupta}},\ }\href{http://link.aps.org/doi/10.1103/PhysRevB.62.R16219}
  {\bibfield  {journal} {\bibinfo  {journal} {Phys. Rev. B}\ }\textbf {\bibinfo
  {volume} {62}},\ \bibinfo {pages} {R16219} (\bibinfo {year}
  {2000})}\BibitemShut {NoStop}%
\bibitem [{\citenamefont {Andersen}\ \emph {\textit{et~al.}}(2000)\citenamefont
  {Andersen}, \citenamefont {Saha-Dasgupta}, \citenamefont {Tank},
  \citenamefont {Arcangeli}, \citenamefont {Jepsen},\ and\ \citenamefont
  {Krier}}]{Andersen2000b}%
  \BibitemOpen
  \bibfield  {author} {\bibinfo {author} {\bibfnamefont {O.~K.}\ \bibnamefont
  {Andersen}}, \bibinfo {author} {\bibfnamefont {T.}~\bibnamefont
  {Saha-Dasgupta}}, \bibinfo {author} {\bibfnamefont {R.~W.}\ \bibnamefont
  {Tank}}, \bibinfo {author} {\bibfnamefont {C.}~\bibnamefont {Arcangeli}},
  \bibinfo {author} {\bibfnamefont {O.}~\bibnamefont {Jepsen}}, \ and\ \bibinfo
  {author} {\bibfnamefont {G.}~\bibnamefont {Krier}},\ }in\ \href@noop {}
  {\emph {\bibinfo {booktitle} {Electronic Structure and Physical Properties of
  Solids. The Uses of the LMTO Method}}},\ \bibinfo {editor} {edited by\
  \bibinfo {editor} {\bibfnamefont {H.}~\bibnamefont {Dreysse}}}\ (\bibinfo
  {publisher} {Springer, New York},\ \bibinfo {year} {2000})\BibitemShut
  {NoStop}%
\bibitem [{\citenamefont {Han}\ \emph {\textit{et~al.}}(2013)\citenamefont
  {Han}, \citenamefont {Wang}, \citenamefont {Marianetti},\ and\ \citenamefont
  {Millis}}]{Han2013}%
  \BibitemOpen
  \bibfield  {author} {\bibinfo {author} {\bibfnamefont {M.~J.}\ \bibnamefont
  {Han}}, \bibinfo {author} {\bibfnamefont {X.}~\bibnamefont {Wang}}, \bibinfo
  {author} {\bibfnamefont {C.~A.}\ \bibnamefont {Marianetti}}, \ and\ \bibinfo
  {author} {\bibfnamefont {A.~J.}\ \bibnamefont {Millis}},\
  }\href{http://link.aps.org/doi/10.1103/PhysRevLett.110.179904} {\bibfield
  {journal} {\bibinfo  {journal} {Phys. Rev. Lett.}\ }\textbf {\bibinfo
  {volume} {110}},\ \bibinfo {pages} {179904} (\bibinfo {year}
  {2013})}\BibitemShut {NoStop}%
\bibitem [{\citenamefont {Haverkort}\ \emph
  {\textit{et~al.}}(2012)\citenamefont {Haverkort}, \citenamefont
  {Zwierzycki},\ and\ \citenamefont {Andersen}}]{Haverkort2012}%
  \BibitemOpen
  \bibfield  {author} {\bibinfo {author} {\bibfnamefont {M.~W.}\ \bibnamefont
  {Haverkort}}, \bibinfo {author} {\bibfnamefont {M.}~\bibnamefont
  {Zwierzycki}}, \ and\ \bibinfo {author} {\bibfnamefont {O.~K.}\ \bibnamefont
  {Andersen}},\ }\href{http://link.aps.org/doi/10.1103/PhysRevB.85.165113}
  {\bibfield  {journal} {\bibinfo  {journal} {Phys. Rev. B}\ }\textbf {\bibinfo
  {volume} {85}},\ \bibinfo {pages} {165113} (\bibinfo {year}
  {2012})}\BibitemShut {NoStop}%
\bibitem [{\citenamefont {Macke}()}]{ReMagX}%
  \BibitemOpen
  \bibfield  {author} {\bibinfo {author} {\bibfnamefont {S.}~\bibnamefont
  {Macke}},\ }\href{http://www.mf.mpg.de/remagx.html} {\bibinfo  {journal}
  {ReMagX x-ray magnetic reflectivity tool - http://remagx.org}\ }\BibitemShut
  {NoStop}%
\bibitem [{\citenamefont {Chantler}(2000)}]{Chantler2000}%
  \BibitemOpen
\bibfield  {journal} {  }\bibfield  {author} {\bibinfo {author} {\bibfnamefont
  {C.~T.}\ \bibnamefont {Chantler}},\
  }\href{http://link.aip.org/link/?JPR/29/597/1} {\bibfield  {journal}
  {\bibinfo  {journal} {J.\ Phys.\ Chem.\ Ref.\ Data}\ }\textbf {\bibinfo
  {volume} {29}},\ \bibinfo {pages} {597} (\bibinfo {year} {2000})}\BibitemShut
  {NoStop}%
\bibitem [{\citenamefont {Shannon}(1976)}]{Shannon1976}%
  \BibitemOpen
  \bibfield  {author} {\bibinfo {author} {\bibfnamefont {R.~D.}\ \bibnamefont
  {Shannon}},\ }\href@noop {} {\bibfield  {journal} {\bibinfo  {journal} {Acta
  Cryst.}\ }\textbf {\bibinfo {volume} {A32}},\ \bibinfo {pages} {751}
  (\bibinfo {year} {1976})}\BibitemShut {NoStop}%
\bibitem [{\citenamefont {Luo}\ and\ \citenamefont {Wang}(2008)}]{Luo2088}%
  \BibitemOpen
  \bibfield  {author} {\bibinfo {author} {\bibfnamefont {X.}~\bibnamefont
  {Luo}}\ and\ \bibinfo {author} {\bibfnamefont {B.}~\bibnamefont {Wang}},\
  }\href{http://link.aip.org/link/?JAP/104/073518/1} {\bibfield  {journal}
  {\bibinfo  {journal} {J.\ Appl.\ Phys.}\ }\textbf {\bibinfo {volume} {104}},\
  \bibinfo {eid} {073518} (\bibinfo {year} {2008})}\BibitemShut {NoStop}%
\bibitem [{\citenamefont {Kareev}\ \emph {\textit{et~al.}}(2008)\citenamefont
  {Kareev}, \citenamefont {Prosandeev}, \citenamefont {Liu}, \citenamefont
  {Gan}, \citenamefont {Kareev}, \citenamefont {Freeland}, \citenamefont
  {Xiao},\ and\ \citenamefont {Chakhalian}}]{Kareev2008}%
  \BibitemOpen
  \bibfield  {author} {\bibinfo {author} {\bibfnamefont {M.}~\bibnamefont
  {Kareev}}, \bibinfo {author} {\bibfnamefont {S.}~\bibnamefont {Prosandeev}},
  \bibinfo {author} {\bibfnamefont {J.}~\bibnamefont {Liu}}, \bibinfo {author}
  {\bibfnamefont {C.}~\bibnamefont {Gan}}, \bibinfo {author} {\bibfnamefont
  {A.}~\bibnamefont {Kareev}}, \bibinfo {author} {\bibfnamefont {J.~W.}\
  \bibnamefont {Freeland}}, \bibinfo {author} {\bibfnamefont {M.}~\bibnamefont
  {Xiao}}, \ and\ \bibinfo {author} {\bibfnamefont {J.}~\bibnamefont
  {Chakhalian}},\ }\href{http://link.aip.org/link/?APL/93/061909/1} {\bibfield
  {journal} {\bibinfo  {journal} {Appl.\ Phys.\ Lett.}\ }\textbf {\bibinfo
  {volume} {93}},\ \bibinfo {eid} {061909} (\bibinfo {year}
  {2008})}\BibitemShut {NoStop}%
\bibitem [{\citenamefont {Kleibeuker}\ \emph
  {\textit{et~al.}}(2010)\citenamefont {Kleibeuker}, \citenamefont {Koster},
  \citenamefont {Siemons}, \citenamefont {Dubbink}, \citenamefont {Kuiper},
  \citenamefont {Blok}, \citenamefont {Yang}, \citenamefont {Ravichandran},
  \citenamefont {Ramesh}, \citenamefont {ten Elshof}, \citenamefont {Blank},\
  and\ \citenamefont {Rijnders}}]{Kleibeuker2010}%
  \BibitemOpen
  \bibfield  {author} {\bibinfo {author} {\bibfnamefont {J.~E.}\ \bibnamefont
  {Kleibeuker}}, \bibinfo {author} {\bibfnamefont {G.}~\bibnamefont {Koster}},
  \bibinfo {author} {\bibfnamefont {W.}~\bibnamefont {Siemons}}, \bibinfo
  {author} {\bibfnamefont {D.}~\bibnamefont {Dubbink}}, \bibinfo {author}
  {\bibfnamefont {B.}~\bibnamefont {Kuiper}}, \bibinfo {author} {\bibfnamefont
  {J.~L.}\ \bibnamefont {Blok}}, \bibinfo {author} {\bibfnamefont {C.-H.}\
  \bibnamefont {Yang}}, \bibinfo {author} {\bibfnamefont {J.}~\bibnamefont
  {Ravichandran}}, \bibinfo {author} {\bibfnamefont {R.}~\bibnamefont
  {Ramesh}}, \bibinfo {author} {\bibfnamefont {J.~E.}\ \bibnamefont {ten
  Elshof}}, \bibinfo {author} {\bibfnamefont {D.~H.~A.}\ \bibnamefont {Blank}},
  \ and\ \bibinfo {author} {\bibfnamefont {G.}~\bibnamefont {Rijnders}},\
  }\href{http://dx.doi.org/10.1002/adfm.201000889} {\bibfield  {journal}
  {\bibinfo  {journal} {Adv.\ Funct.\ Mater.}\ }\textbf {\bibinfo {volume}
  {20}},\ \bibinfo {pages} {3490} (\bibinfo {year} {2010})}\BibitemShut
  {NoStop}%
\end{thebibliography}
\end{document}